\documentclass[11pt,reqno]{amsart}
\usepackage{amsfonts,amsmath}
\usepackage{graphicx}
\usepackage{epstopdf}
\vspace{4ex}
\begin{document}
\vspace*{1cm}
\begin{center}
{\Large \bf Generalised discriminants, deformed quantum
Calogero-Moser system and Jack polynomials}
\\*[4ex]
{\bf A.N. Sergeev$^{\dagger}$ and A.P. Veselov$^{\ddagger,\star}$}
\\*[2ex]
\end{center}

\noindent $^\dagger$ Balakovo Institute of Technology and
Control, Balakovo,413800, Russia

\noindent $^\ddagger$
Department of Mathematical Sciences, Loughborough University,\\
Loughborough,  LE11 3TU, UK

\noindent $^\star$ Landau Institute
for Theoretical Physics, Kosygina 2, Moscow, 117940, Russia

\noindent E-mail addresses: sergeev@bittu.org.ru,
A.P.Veselov@lboro.ac.uk

\vspace*{1cm}

{\small  {\bf Abstract.} It is shown that the deformed
Calogero-Moser-Sutherland (CMS) operators can be described as the restrictions
on certain affine subvarieties (called generalised discriminants) of the usual CMS
operators for infinite number of particles. The ideals of these varieties are shown
to be generated by the Jack symmetric functions related to the Young diagrams with special geometry.
A general structure of the ideals which are invariant under the action of the quantum CMS integrals
is discussed in this context. The shifted super-Jack polynomials are introduced and
combinatorial formulas for them and for super-Jack polynomials are given.}

\section{Introduction}

The primary goal of this paper is to explain the algebraic nature
of integrability of the deformed Calogero-Moser-Sutherland (CMS)
operators
\begin{eqnarray}
\label{anm} L_{n,m,\theta} = -\left (\frac{\partial^2}{{\partial
x_{1}}^2}+\dots +\frac{\partial^2}{{\partial x_{n}}^2}\right)-
k\left(\frac{\partial^2}{{\partial y_{1}}^2}+\dots
+\frac{\partial^2}{{\partial y_{m}}^2}\right) \nonumber \\
+\sum_{i<j}^{n}\frac{2k(k+1)}{\sin^2(x_{i}-x_{j})}
+\sum_{i<j}^{m}\frac{2(k^{-1}+1)}{\sin^2(y_{i}-y_{j})}
+\sum_{i=1}^{n}\sum_{j=1}^{m}\frac{2(k+1)}{\sin^2(x_{i}-y_{j})}
\end{eqnarray}
For $m=1$ these operators have been first introduced in \cite{CFV,
CFV1}, for general $m$ they were considered in \cite{Ser, Ser1}.
In \cite{SV} we have suggested a general construction of the
deformed CMS operators related to Lie superalgebras, which in the
case of Lie superalgebra $sl(n|m)$ leads to the operators
(\ref{anm}). Unfortunately this relation with Lie superalgebras
itself does not supply the integrability of this problem, which
was proved in \cite{SV} by direct construction of the quantum
integrals.

In this paper we will present a more conceptual proof of the integrability
of (\ref{anm}) by showing that (after some gauge transformation
and change of variables) the deformed CMS operators can be
described as the {\it restriction} of the usual CMS operators
for infinite number of particles onto certain subvarieties of Macdonald variety
called generalised discriminants.  Note that the restriction of a
differential operator onto a submanifold is possible only under
very special circumstances. In case of the algebraic subvariety
this means that the corresponding ideal must be invariant under
the action of the operator.

For the proof we use the theory of Jack polynomials \cite{Ma}, \cite{Stanley} and
the theory of shifted Jack polynomials developed recently by Knop,
Sahi, Okounkov and Olshanski \cite{Sa, KSa, OkO}. We have been
partially inspired by a recent very interesting paper
\cite{FJMM} by B. Feigin, Jimbo, Miwa and Mukhin, where certain
ideals in the rings of symmetric polynomials were described in
terms of Jack polynomials.

The structure of the paper is following. First we review the basic facts
from the theory of Jack and shifted Jack
polynomials and from the theory of Cherednik-Dunkl operators.

In section 5 we introduce the generalised
discriminants and prove our main result. In section 6 we show
that the quantum integrals we have constructed in \cite{SV} can also be described as
the restrictions of certain integrals of the usual CMS problem. The notion of the
shifted super-Jack polynomials naturally appears in this relation.

Section 7 is devoted to the description of the ideals in the algebra
of symmetric functions which are invariant under the action of the
quantum integrals of the CMS system. We show that the rectangular
Young diagrams related to generalised discriminants play a very special role here.

In the last section we give combinatorial formulas for the super-Jack and shifted
super-Jack polynomials generalising the results of Stanley, Okounkov and Olshanski
\cite{Stanley,OkO,Ok2}.

\section{ Symmetric functions and Jack polynomials}

In this section we recall some general facts about symmetric
functions and Jack polynomials mainly following Macdonald's book
\cite{Ma}. It would be convenient for us to use instead of the
parameter $\alpha$ in Macdonald's notations of Jack polynomials
the parameter $$\theta = \frac{1}{\alpha}$$ (cf. \cite{OkO}). It
is different from the parameter $k$ used in our previous work
\cite{SV} by sign change: $$\theta = -k.$$

Let $P_{N}={\mathbb C}[x_{1},\dots,x_{N}]$ be the polynomial
algebra in $N$ independent variables and $\Lambda_{N}\subset
P_{N}$ be the subalgebra of symmetric polynomials.

A {\it partition} is any sequence $$
\lambda=(\lambda_{1},\lambda_{2},\dots,\lambda_{r}\,\dots) $$ of
nonnegative integers in decreasing order $$
\lambda_{1}\ge\lambda_{2}\ge\dots\ge\lambda_{r}\,\ge\dots $$
containing only finitely many nonzero terms. The number of nonzero
terms in $\lambda$ is the {\it length} of $\lambda$ denoted by
$l(\lambda)$. The sum $\mid\lambda\mid =
\lambda_{1}+\lambda_{2}+\dots$ is called the {\it weight} of
$\lambda$. The set of all partitions of weight $N$ is denoted by
${\mathcal P}_N.$

On this set there is a natural involution: in the standard
diagrammatic representation \cite{Ma} it corresponds to the
transposition (reflection in the main diagonal). The image of a
partition $\lambda$ under this involution is called the {\it
conjugate} of $\lambda$ and denoted by $\lambda'.$ This involution
will play an essential role in our paper.

Partitions can be used to label the bases in the symmetric algebra
$\Lambda_{N}.$ There are several important bases in $\Lambda_{N}$.

1) {\it Monomial symmetric polynomials} $m_{\lambda}, \lambda \in
{\mathcal P}_N$ are defined by $$
m_{\lambda}(x_{1},\dots,x_{N})=\sum
x_{1}^{a_{1}}x_{2}^{a_{2}}\dots x_{N}^{a_{N}} $$ summed over all
distinct permutations $a$ of $\lambda =
(\lambda_{1},\lambda_{2},\dots,\lambda_{N}).$

2) {\it Elementary symmetric polynomials} are defined by $$
\sum_{k=o}^{\infty} e_{k}t^k=\prod_{i\ge 1}(1+x_{i}t).$$ For each
partition $\lambda$ we define $$
e_{\lambda}=e_{\lambda_{1}}e_{\lambda_{2}}\dots $$

3) Similarly {\it complete symmetric polynomials} are defined by
$$ \sum_{k=o}^{\infty} h_{k}t^k=\prod_{i\ge 1}(1-x_{i}t)^{-1} $$
and $$ h_{\lambda}=h_{\lambda_{1}}h_{\lambda_{2}}\dots.$$

4) Finally the most convenient for us will be the {\it power sums}
$$ p_{k}=x_{1}^{k}+x_{2}^{k}+\dots ,$$ where again for any
partition $\lambda$ $$
p_{\lambda}=p_{\lambda_{1}}p_{\lambda_{2}}\dots $$

It is well-known \cite{Ma} that each of these sets of functions
with $l(\lambda)\le N$ form a basis in $\Lambda_{N}.$

We will need the following infinite dimensional versions
of both $P_{N}$ and $\Lambda_{N}$. Let $M\le N$ and $\varphi_{N,M}
: P_{N}\longrightarrow P_{M}$ be the homomorphism which sends each
of $x_{M+1},\dots,x_{N}$ to zero and other $x_{i}$ to themselves.
It is clear that $\varphi_{N,M} (\Lambda_{N})=\Lambda_{M}$ so we
can consider the inverse limits in the category of graded algebras
$$ P=\lim_{\longleftarrow}
P_{N},\quad\Lambda=\lim_{\longleftarrow} \Lambda_{N}.$$ This means
that $$ P=\oplus_{r=0}^{\infty} P^{r},\quad
P^{r}=\lim_{\longleftarrow}P_{N}^{r} $$ $$
\Lambda=\oplus_{r=0}^{\infty}
\Lambda^{r},\quad\Lambda^{r}=\lim_{\longleftarrow}\Lambda_{N}^{r}
$$ where $P_{N}^r, \Lambda_{N}^r$ are the homogeneous components
of $P_{N}, \Lambda_{N}$ of degree $r$. The elements of $\Lambda$
are called {\it symmetric functions.}

Since for any partition $\lambda$  $$
\varphi_{N,M}(m_{\lambda}(x_{1},\dots,x_{N}))=m_{\lambda}(x_{1},\dots,x_{M})$$
(and similarly for the polynomials $h, e, p$) we can define the
symmetric functions $m_{\lambda}, h_{\lambda}, e_{\lambda},
p_{\lambda}.$

Another important example of symmetric functions are {\it Jack
polynomials}. We give here their definition in the form most
suitable for us.

Recall that on the set of partitions ${\mathcal P}_N$ there is the
following {\it dominance partial ordering}: we write
$\mu\le\lambda$ if for all $i \geq 1$ $$
\mu_{1}+\mu_{2}+\dots+\mu_{i}\le\lambda_{1}+\lambda_{2}+\dots+\lambda_{i}.
$$

Consider the following {\it CMS operator}
\begin{equation}
\label{CMS}
 {\mathcal L}_{\theta}^{(N)}=\sum_{i=1}^N
\left(x_{i}\frac{\partial}{\partial
x_{i}}\right)^2+\theta\sum_{1\le i < j\le N}
\frac{x_{i}+x_{j}}{x_{i}-x_{j}}\left(
x_{i}\frac{\partial}{\partial x_{i}}-
x_{j}\frac{\partial}{\partial
x_{j}}\right)-\theta(N-1)\sum_{i=1}^N
x_{i}\frac{\partial}{\partial x_{i}}$$ $$ = \sum_{i=1}^N
\left(x_{i}\frac{\partial}{\partial x_{i}}\right)^2+2\theta\sum_{
i \ne j} \frac{x_{i}x_{j}}{x_{i}-x_{j}}\frac{\partial}{\partial
x_{i}}.
\end{equation}
It coincides with the standard (trigonometric) Calogero-Moser-Sutherland operator
\cite{C,Suth} if we change the gauge and use exponential
coordinates. It is related to the so-called Laplace-Beltrami
operator $$ {\Box}_{N}^{\alpha}=\frac{\alpha}{2}\sum_{i=1}^N
x_{i}^2\left(\frac{\partial}{\partial x_{i}}\right)^2+\sum_{i\ne
j}\frac{x_{i}^2}{x_{i}-x_{j}}\frac{\partial}{\partial
x_{i}}-\sum_{i=1}^N x_{i}\frac{\partial}{\partial x_{i}} $$ used
by Macdonald \cite{Ma} by a simple formula $$ {\mathcal
L}_{\theta}^{(N)}=
2\theta{\Box}_{N}^{\frac{1}{\theta}}-\sum_{i=1}^N
x_{i}\frac{\partial}{\partial x_{i}}.$$

An important property of the CMS operator is its {\it stability}
under the change of $N:$ the following diagram is commutative
$$
\begin{array}{ccc}
\Lambda_{N}&\stackrel{{\mathcal
L}_{\theta}^{(N)}}{\longrightarrow}&\Lambda_{N} \\ \downarrow
\lefteqn{\varphi_{N,M}}& &\downarrow \lefteqn{\varphi_{N,M}}\\
\Lambda_{M}&\stackrel{{\mathcal
L}_{\theta}^{(M)}}{\longrightarrow}&\Lambda_{M} \\
\end{array}
$$ (see \cite {Ma}, example 3 on the page 326). This allows us to
define the CMS operator ${\mathcal L}_{\theta}$ on the space of
symmetric functions $\Lambda$ as the limit of ${\mathcal
L}_{\theta}^{(N)}.$

{\bf Theorem-Definition.} {\it If $\theta$ is not negative
rational number or zero than for  any partition $\lambda$  ,
$l(\lambda)\le N$ there is a unique polynomial
$P_{\lambda}(x,\theta) \in \Lambda_{N}$ (called Jack polynomial)
such that

1) $P_{\lambda}(x,\theta)= m_{\lambda}+\sum_{\mu<\lambda}
u_{\lambda\mu}m_{\mu}$, where $u_{\lambda\mu}\in{\mathbb C}$

2) $ P_{\lambda}(x,\theta)$ is an eigenfunction of the CMS
operator ${\mathcal L}_{\theta}^{(N)}.$}

To prove this one can check that the operator ${\mathcal
L}^{(N)}_{\theta}$ has an upper triangular matrix in the monomial basis
$m_{\mu}$: $$ {\mathcal
L}_{\theta}^{(N)}(m_{\lambda})=\sum_{\mu\le\lambda}
c_{\lambda\mu}m_{\mu} $$ where the coefficients $c_{\lambda\mu}$
can be described explicitly (see \cite{Ma}, page 327 and
\cite{LLM}). In particular $$
c_{\lambda\lambda}=\sum_{i=1}^N\lambda_{i}^2-2\theta\sum_{i=1}^N
(i-1)\lambda_{i}=2n(\lambda^{\prime})-2\theta n(\lambda)+|\lambda|
$$ where $n(\lambda)=\sum_{i\ge 1}(i-1)\lambda_{i}$. It is easy to
see that if  $\lambda > \mu$ then $n(\lambda)<n(\mu)$ and
$n(\lambda^{\prime})>n(\mu^{\prime})$. So we have $$
c_{\lambda\lambda} -
c_{\mu\mu}=2[n(\lambda)-n(\mu)]+2\theta[n(\mu^{\prime})-n(\lambda^{\prime})]
$$ Since $\theta$ is not a negative rational we see that $ c_{\lambda\lambda}
\ne c_{\mu\mu}$ if $\lambda
> \mu$. This implies the claim.

From the stability of the CMS operators it follows that
$$\varphi_{N,M}(P_{\lambda}(x_{1},\dots,x_{N}))=P_{\lambda}(x_{1},\dots,x_{M})
$$ so we have correctly defined Jack symmetric functions
$P_{\lambda}(x,\theta)\in \Lambda$ which are the eigenfunctions of
the CMS operator ${\mathcal L}_{\theta}.$

\section{Shifted symmetric functions and shifted Jack polynomials.}

We discuss now the so-called {\it shifted Jack polynomials}
introduced recently by Knop, Sahi, Okounkov and Olshanski
\cite{Sa,KSa,OkO,Ok2}. For a nice review of the
theory of shifted symmetric functions and its role in the
representation theory of the symmetric group we refer to
\cite{IO}.

Let us denote by $\Lambda _{N,\theta} $ the algebra of polynomials
$f(x_{1},\dots,x_{N})$ which are symmetric in the shifted variables $
x_{i}+\theta(1-i)$. This algebra has the filtration by the degree
of polynomials: $$ (\Lambda_{N,\theta})_{0}\subset
(\Lambda_{N,\theta})_{1}\subset \dots\subset
(\Lambda_{N,\theta})_{r}\subset\dots $$ We have the following
shifted analog of power sums:
\begin{equation}
\label{ps} p^*_{r}(x_{1},\dots,x_{N},\theta)=\sum_{i=1}^N
\left[(x_{i}+\theta(1-i))^r-(\theta(1-i))^r\right].
\end{equation}
The polynomials $$
p^*_{\lambda}(x,\theta)=p^*_{\lambda_{1}}(x,\theta)p^*_{\lambda_{2}}(x,\theta)\dots
$$ where
$\lambda=(\lambda_{1},\lambda_{2},\dots,\lambda_{r}\,\dots)$ is a
partition of $N$ form a basis in $\Lambda _{N,\theta}.$ Since
$\varphi_{N,M} (\Lambda_{N,\theta})=\Lambda_{M,\theta}$ one can
consider the inverse limit $$\Lambda_{\theta}
=\lim_{\longleftarrow} \Lambda_{N,\theta}$$ in the category of
filtered algebras: $$ \Lambda_{\theta}=\bigcup_{r=0}^{\infty}
(\Lambda_{\theta})_{r},\quad(\Lambda_{\theta})_{r}=\lim_{\longleftarrow}(\Lambda_{N,\theta})_{r}.
$$ The algebra $\Lambda_{\theta}$ is called the algebra of {\it
shifted symmetric functions.}

The {\it shifted Jack polynomials} can be defined in the following
way (see \cite{KSa},\cite{OkO}). Let us introduce the
following function on the set of partitions
\begin{equation}
\label{H}
 H(\lambda,\theta)=\prod_{\Box\in\lambda}(
c_{\theta}^{'}(\Box)+1) \end{equation} Here we use the
diagrammatic representation of the partitions using the squares
\cite{Ma} and to each square $\Box = (i,j)$ we prescribe the numbers
\begin{equation}
\label{c'}
c_{\theta}^{'}(\Box)=\lambda_{i}-j+\theta(\lambda^{'}_{j}-i).
\end{equation}

{\bf Theorem-Definition \cite{KSa, OkO}.} {\it Let $\lambda$ be a
partition with $\lambda_{N+1}=0$. There exists a unique shifted
symmetric polynomial $P_{\lambda}^{*}(x,\theta) \in
\Lambda_{N,\theta}$ (called shifted Jack polynomial) such that
$\deg P_{\lambda}\le |\lambda|$ and $$
P_{\lambda}^{*}(\mu,\theta)=\left\{ \begin{aligned}
\quad H(\lambda),& \quad\quad \quad\mu=\lambda\\
 \quad 0, &\quad \quad |\mu|\le|\lambda|,\mu\ne\lambda,\mu_{N+1}=0
 \end{aligned}
 \right. $$}

Here and later throughout the paper by $P(\lambda)$ for a
polynomial $P(x_1,\dots,x_N)$ and a partition
$\lambda=(\lambda_{1},\dots,\lambda_{N})$ we mean
$P(\lambda_1,\dots,\lambda_N).$ It is actually a very useful idea
to view the shifted symmetric polynomials as the functions on the
partitions (see e.g. \cite{IO}).

It is easy to see from the definition that the shifted Jack
polynomials are stable: $$
\varphi_{N,M}(P_{\lambda}^{*}(x_{1},\dots,x_{N},\theta))=P_{\lambda}^{*}(x_{1},\dots,x_{M},\theta),
$$ so we can define the shifted Jack function
$P_{\lambda}^{*}(x,\theta)$ as an element of $\Lambda_{\theta}.$

Knop and Sahi \cite{KSa} proved that the shifted Jack polynomials
satisfy the following so-called {\it Extra Vanishing Property}:
\begin{equation}
\label{EVP} P^*_{\lambda}(\mu,\theta) = 0
\end{equation}
unless the diagram of $\lambda$ is a subset of the diagram of
$\mu$ and that $P_{\lambda}^{*}(x,\theta)$ is the usual Jack
polynomial $P_{\lambda}(x,\theta)$ plus lower order terms.

We will need the following {\it duality} property of the shifted Jack
polynomials:
\begin{equation}
\label{dual}
P_{\lambda}^{*}(\mu',\theta)=\frac{H(\lambda,\theta)}{H(\lambda^{\prime},1/\theta)}P_{\lambda'}^{*}(\mu,1/\theta).
\end{equation}
For shifted Macdonald polynomials this was proved by Okounkov \cite{Ok2}.
We present here an independent proof for Jack polynomials based
on the Bernoulli sums.

Consider the following natural {\it
conjugation homomorphism}:
\begin{equation}
\label{omega} (\omega^*(f))(\lambda)=f(\lambda^{\prime})
\end{equation}

{\bf Proposition 1.} {\it The conjugation homomorphism maps the
algebra of shifted symmetric functions $\Lambda_{\theta}$ into the
algebra $\Lambda_{1/\theta}.$}

{\bf Proof.} Let us introduce the following {\it Bernoulli sums}:
\begin{equation}
\label{Bernoulli}
 b_{k}(x_{1},\dots,x_{N},\theta)=\sum_{i=1}^N
\left[B_{k}(x_{i}+\theta(1-i))-B_{k}(\theta(1-i))\right],
\end{equation}
where $B_{k}(x)$ are the classical Bernoulli polynomials. Recall that
$B_{k}(x)$ can be defined through the generating function $$
\frac{te^{xt}}{e^t-1}=\sum_{k=0}^{\infty}B_{k}(x)\frac{t^k}{k!} $$
and satisfy the property $$ B_{k}(x+1)-B_{k}(x)=kx^{k-1} $$ or,
more generally
\begin{equation}
\label{Bernou}
 B_{k}(x+l)-B_{k}(x)=k\sum_{i=1}^l (x+i-1)^{k-1}.
\end{equation}
It is easy to see that the Bernoulli sums are stable: $$
\varphi_{N,M}(b_{k}(x_{1},\dots,x_{N},\theta))=b_{k}(x_{1},\dots,x_{M},\theta),
$$ so one can define the {\it shifted symmetric Bernoulli functions}
$b_{k}(x,\theta) \in \Lambda_{\theta}$.

{\bf  Lemma 1.}  {\it Bernoulli functions satisfy the following
symmetry:}\begin{equation} \label{Berno}
b_{k}(\lambda^{\prime},\theta)=(-\theta)^{k-1}b_{k}(\lambda,1/\theta).
\end{equation}

The proof is a straightforward check: $$
b_{k}(\lambda^{\prime},\theta)=\sum_{j=1}^{l(\lambda^{\prime})}\left[B_{k}(\lambda_{j}^{\prime}+\theta(1-j))-B_{k}(\theta(1-j))\right]=
$$ $$
k\sum_{j=1}^{l(\lambda^{\prime})}\sum_{i=1}^{\lambda_{j}^{\prime}}\left[\theta(1-j)+i-1\right]^{k-1}=
$$ $$ k\sum_{\Box\in\lambda}\left[\theta(1-j)+i-1\right]^{k-1}
=k(-\theta)^{k-1}\sum_{\Box\in\lambda}\left[1/\theta(1-i)+
j-1\right]^{k-1} $$ $$
k(-\theta)^{k-1}\sum_{i=1}^{l(\lambda)}\sum_{i=1}^{\lambda_{i}}\left[1/\theta(1-i)+
j-1\right]^{k-1}
=
$$ $$
(-\theta)^{k-1}\sum_{i=1}^{l(\lambda)}\left[B_{k}(\lambda_{i}+1/\theta(1-i))-B_{k}(\theta(1-i))\right]=(-\theta)^{k-1}b_{k}(\lambda,1/\theta)
$$ We have used here the property (\ref{Bernou}) of the Bernoulli
polynomials.

Since Bernoulli sums $b_k(x, \theta)$ generate the algebra
$\Lambda_{\theta}$ the proposition now follows. Combining this
with the definition of the shifted Jack polynomials we have the
duality property (\ref{dual}).

\section{Cherednik - Dunkl operators and Harish-Chandra homomorphism}

In this section we present the basic facts about Cherednik-Dunkl
operators. For the details we refer to Opdam's review
\cite{Opdam}.

By {\it Cherednik - Dunkl operators} we mean the following
difference-differential operators
\begin{equation}
\label{CD}
 D_{i,N} = x_{i}\frac{\partial}{\partial
x_{i}}+\theta\sum_{j\ne i} \frac{x_{\max \{i,j
\}}}{x_{i}-x_{j}}(1-\sigma_{ij}), \quad i,j \leq N,
\end{equation}
where $\sigma_{ij}$ is acting on the function $\phi(x_1,\dots
x_N)$ by permutation of $i$-th and $j$-th coordinates. If $i > N$ we assume that
$D_{i,N}=0.$ We should warn the reader that our
definition is different from the usual one $$\hat D_{i,N} =
x_{i}\frac{\partial}{\partial x_{i}}+\theta\sum_{j\ne i}
\frac{x_{\max \{i,j \}}}{x_{i}-x_{j}}(1-\sigma_{ij}) +
\theta(N-i),$$ (see e.g. \cite{Opdam}) by a shift. This shift is
necessary for the stability of $D_{i,N}$ and explains the relations with the theory of shifted symmetric
functions which we discuss below.

The first important property of the Cherednik - Dunkl operators is
that they commute with each other: $$[D_{i,N}, D_{j,N}] = 0.$$
This means that one can substitute them in any polynomial $P$ in
$N$ variables without ordering problems.

The second property is that if one does this for a shifted
symmetric polynomial $f \in \Lambda_{N,\theta}$ then the
corresponding operator $f(D_{1,N} \dots D_{N,N})$ leaves the
algebra of symmetric polynomials $\Lambda_N$ invariant: $$
f(D_{1,N} \dots D_{N,N}): \Lambda_N \rightarrow \Lambda_N.$$ The
restriction of the operator $f(D_{1,N} \dots D_{N,N})$ on the
algebra $\Lambda_N$ is given by some differential operator, which
we will denote as ${\mathcal L}^f_{N,\theta}.$ A formula for this operator
can be found by moving all the permutation operators in $f(D_{1,N}
\dots D_{N,N})$ to the right using natural commutation relations
and then erasing them.

One can check that if we apply this operation to the shifted
square sum $p^*_{2}(x_{1},\dots,x_{N},\theta)=\sum_{i=1}^N
\left[(x_{i}+\theta(1-i))^2-(\theta(1-i))^2\right]$ we arrive at
the CMS operator (\ref{CMS}). Thus all the operators ${\mathcal
L}^f_{N, \theta}$ are actually the quantum integrals of the CMS problem. The
Jack polynomials are the joint eigenfunctions of all these
operators: if $P_{\lambda}(x,\theta)$ is the Jack polynomial
corresponding to a partition $\lambda$ of length $N$ then
\begin{equation} \label{LP}
{\mathcal L} ^f_{N, \theta} P_{\lambda}(x,\theta )=
f(\lambda_{1},\lambda_{2},\dots,\lambda_{N})P_{\lambda}(x,\theta.)
\end{equation}

This allows us to define a homomorphism (which is actually a
monomorphism) $\chi: f \rightarrow {\mathcal L}^f_{N, \theta}$ from the
algebra $\Lambda_{N, \theta}$ to the algebra of differential
operators. Let us denote by ${\mathcal D}(N,\theta)$ the image of
$\chi$. The inverse homomorphism $$ \chi^{-1} : {\mathcal
D}(N,\theta)\longrightarrow \Lambda _{N,\theta} $$ is called the
{\it Harish-Chandra isomorphism.} It can be defined by the action
on the Jack polynomials: the image of ${\mathcal L} \in {\mathcal
D}(N,\theta)$ is a polynomial $f = f_{\mathcal L} \in \Lambda
_{N,\theta}$ such that $${\mathcal L} P_{\lambda}(x,\theta)=
f(\lambda)P_{\lambda}(x,\theta).$$

One can check that the Cherednik-Dunkl operators $D_{i,N}$ are stable (note that this is not true for the shifted
operators $\hat D_{i,N}$): the diagram $$
\begin{array}{ccc}
P_{N}&\stackrel{D_{i,N}}{\longrightarrow}&P_{N} \\ \downarrow
\lefteqn{\varphi_{N,M}}& &\downarrow \lefteqn{\varphi_{N,M}}\\
P_{M}&\stackrel{D_{i,M}}{\longrightarrow}&P_{M} \\
\end{array}
$$ is commutative for all $M\le N$ and $i\ge 1$. Similarly for any
$f\in \Lambda_{N,\theta}$ and $g=\varphi_{N,M}(f), M \le N$ the
following diagram is commutative: $$
\begin{array}{ccc}
\Lambda_{N}&\stackrel{{\mathcal
L}^f_{N, \theta}}{\longrightarrow}&\Lambda_{N} \\ \downarrow
\lefteqn{\varphi_{N,M}}& &\downarrow \lefteqn{\varphi_{N,M}}\\
\Lambda_{M}&\stackrel{{\mathcal
L}^g_{M, \theta}}{\longrightarrow}&\Lambda_{M} \\
\end{array}
$$ This allows us to define for any shifted symmetric function
$f\in \Lambda_{\theta}$ a differential operator $${\mathcal L}^{f}_{\theta}
: \Lambda \longrightarrow \Lambda$$ and the infinite dimensional
version of the homomorphism $\chi.$ We will denote by ${\mathcal
D} (\theta)$ the image of this homomorphism, which can be
interpreted as the ring of quantum integrals of the
(infinite-dimensional) CMS problem. The inverse (Harish-Chandra)
homomorphism $\chi^{-1}: {\mathcal D} (\theta) \longrightarrow
\Lambda_{\theta}$ can be described by the relation $${\mathcal
L}^{f}_{\theta} P(\lambda, \theta) = f(\lambda) P(\lambda, \theta),$$ where
now $f \in \Lambda_{\theta}$ and $P(\lambda, \theta)$ is Jack
symmetric function.

\section{Generalised discriminants and deformed CMS operators.}

Now we are ready to formulate our main results. The central role
in our construction will play the following algebra
$\Lambda_{n,m,\theta}$ introduced in \cite{SV} (denoted there as $\Lambda^{0}_{n,m,\theta} $).

Let $P_{n,m}={\mathbb C}[x_{1},\dots,x_{n},y_{1},\dots,y_{m}]$ be
the polynomial algebra in $n+m$ independent variables. Then
$\Lambda_{n,m,\theta}\subset P_{n,m}$ is the subalgebra consisting
of polynomials which are symmetric in $x_{1},\dots,x_{n}$ and
$y_{1},\dots,y_{m}$ separately and satisfy the conditions
\begin{equation}
\label{cond0}
 \left(x_{i}\frac{\partial}{\partial
x_{i}}+\theta y_{j}\frac{\partial}{\partial y_{j}}\right)f\equiv 0
\end{equation}
or, equivalently
\begin{equation} \label{cond0*}
 \left(\frac{\partial}{\partial
x_{i}}+\theta \frac{\partial}{\partial y_{j}}\right)f\equiv 0
\end{equation}
on each hyperplane  $x_{i}-y_{j}=0$ for $i=1,\dots,n$ and
$j=1,\dots,m.$ It is shown in \cite{SV} that for generic $\theta$
(namely, if $\theta$ is not a negative rational or zero)
$\Lambda_{n,m,\theta}$ coincides with its subalgebra ${\mathcal
N}_{n,m,\theta}$ generated by the {\it deformed Newton sums}
\begin{equation}
\label{defNewton}
 p_{r}(x,y,\theta)= \sum_{i=1}^n
{x_{i}^r}-\frac{1}{\theta}\sum_{j=1}^m {y_{j}^r}
\end{equation}
which obviously belong to $\Lambda_{n,m,\theta}$ for all
nonnegative integers $r$.

This algebra has appeared in \cite{SV} in relation with the following {\it
deformed CMS operator}
\begin{equation}
\label{defCMS}
 {\mathcal L}_{n,m,\theta}=\sum_{i=1}^n
\left(x_{i}\frac{\partial}{\partial
x_{i}}\right)^2-\theta\sum_{j=1}^m
\left(y_{j}\frac{\partial}{\partial
y_{j}}\right)^2+\theta\sum_{1\le i < j\le n}
\frac{x_{i}+x_{j}}{x_{i}-x_{j}}\left(
x_{i}\frac{\partial}{\partial x_{i}}-
x_{j}\frac{\partial}{\partial x_{j}}\right)- $$ $$ \sum_{1\le i <
j\le m} \frac{y_{i}+y_{j}}{y_{i}-y_{j}}\left(
y_{i}\frac{\partial}{\partial y_{i}}-
y_{j}\frac{\partial}{\partial y_{j}}\right)-\sum_{i=1 }^n\sum_{
j=1}^ m \frac{x_{i}+y_{j}}{x_{i}-y_{j}}\left(
x_{i}\frac{\partial}{\partial x_{i}}+\theta
y_{j}\frac{\partial}{\partial y_{j}}\right)- $$ $$
(\theta(n-1)-m)\left(\sum_{i=1}^n x_{i}\frac{\partial}{\partial
x_{i}}+\sum_{j=1}^m y_{j}\frac{\partial}{\partial x_{j}}\right)
\end{equation}

{\bf  Lemma 2.}  {\it The deformed CMS operator preserves the
algebra} $\Lambda_{n,m,\theta}$:
\begin{equation} \label{invar}
{\mathcal L}_{n,m,\theta}: \Lambda_{n,m,\theta} \rightarrow
\Lambda_{n,m,\theta}
\end{equation}

The proof follows from a more general statement proved in the next
section (see Theorem 5).

Let now $\Lambda$ be the algebra of symmetric functions in
infinite number of variables $z_{1},z_{2},\dots$,
$p_{r}(z)=z_{1}^r+z_{2}^r+\dots$ be the power sums,
$P_{\lambda}(z,\theta)$ be the Jack polynomials (see Section 2 above).

Consider the following homomorphism $\varphi$ from $\Lambda$ to
$\Lambda_{n,m,\theta}$ such, that $$ \varphi (p_{r}(z))=
p_{r}(x,y,\theta). $$ Since $p_r(z)$ are free generators of $\Lambda$ this determines
$\varphi$ uniquely. Such a homomorphism was first used by Kerov,
Okounkov and Olshanski in \cite{KOO}.

Our central result can be formulated as follows. Let ${\mathcal
L}_{\theta}$ be the usual CMS operator in infinite dimension.

{\bf Theorem 1.}  {\it The following diagram is commutative for
all values of the parameter} $\theta$:
\begin{equation} \label{commdia}
\begin{array}{ccc}
\Lambda&\stackrel{{\mathcal L}_{\theta}}{\longrightarrow}&\Lambda
\\ \downarrow \lefteqn{\varphi}& &\downarrow \lefteqn{\varphi}\\
\Lambda_{n,m, \theta}&\stackrel{{\mathcal
L}_{n,m,\theta}}{\longrightarrow}&\Lambda_{n,m,\theta} \\
\end{array}
\end{equation}

Before going to the proof let us discuss the geometric meaning of
this claim. For generic values of the parameter $\theta$ according 
to \cite{SV} the algebra $\Lambda_{n,m,\theta}$ is finitely generated, 
so we can introduce an affine algebraic variety $${\mathcal D}_{n,m,\theta} = Spec \,
\Lambda_{n,m,\theta} = Spec \,{\mathcal N}_{n,m,\theta}.$$ We
call this variety {\it generalised discriminant} because for
special values $m=1$ and $\theta = -1/2$ the algebra ${\mathcal
N}_{n,1, -1/2}$ coincides with the algebra of functions on the
standard discriminant variety ${\mathcal D}_{n+1}$ of the
polynomials of degree $n+1$ having a multiple root.
A more general case $\theta = -1/q$ corresponds to the stratum in the discriminant variety 
when one of the roots has a multiplicity at least $q,$
so informally speaking the generalised discriminant ${\mathcal D}_{n,m,\theta} $
consists of "polynomials" with $n$ simple roots and $m$ roots of "multiplicity" $\mu = -\theta^{-1}$.

We understand that our terminology is not perfect since the term "discriminant"
was used and generalised in many different ways (see e.g. well-known book
\cite{GKZ} by Gelfand, Kapranov and Zelevinsky) but an alternative term "generalised coincident root loci" looks too long and not much better. We would like to mention that the problem of finding the algebraic equations defining the strata in the discriminants is non-trivial and goes back to Arthur Cayley \cite{Cayley} (see \cite{Weyman, Chip, KMSV} for the recent results in this direction).

The map $\varphi$ determines an embedding of ${\mathcal
D}_{n,m,\theta}$ into ${\mathcal M} = Spec \, \Lambda.$ We will
call ${\mathcal M}$ {\it Macdonald variety} although strictly
speaking it is defined only as an affine scheme since the algebra
$\Lambda$ is not finitely generated. Since the algebra
$\Lambda_{n,m,\theta}$ is also not finitely generated for special
positive rational $\theta$ (see \cite{SV}) the Macdonald variety
is a proper space for the generalised discriminants to live in.

Notice that for negative rationals the algebra $\Lambda_{n,m,\theta}$
could be bigger than ${\mathcal N}_{n,m,\theta}$, so in general we should distinguish
the variety ${\mathcal D}_{n,m,\theta} = Spec \,\Lambda_{n,m,\theta} $
and its embedding in ${\mathcal M}$  $\tilde {\mathcal D}_{n,m,\theta} =  Spec \,{\mathcal N}_{n,m,\theta}.$

{\bf Corollary.}  {\it The deformed CMS operator (\ref{defCMS}) is
the restriction of the usual CMS operator ${\mathcal L}_{\theta}$
on Macdonald variety onto the generalised discriminant subvariety}
${\mathcal D}_{n,m,\theta}.$

We should mention that the restriction of a differential operator ${\mathcal
L}$ onto a subvariety is a very rare phenomenon. Indeed this is
possible only when the ideal corresponding to this subvariety is
${\mathcal L}$-invariant. In our case the situation is even more
peculiar because we have a finite-dimensional subvariety in
infinite dimensions.

{\bf Proof of Theorem 1.} Let us introduce the following function
$\Pi \in \Lambda[[t_1, \dots, t_N]]$ which plays an important role
in the theory of Jack polynomials (see \cite{Stanley}):
$$ \Pi = \prod_{l = 1}^{N} \prod_{i \geq l} (1-z_{i}t_{l})^{-\theta}.$$

{\bf Lemma 3}. {\it The function $\Pi$ satisfies the following
properties:}
\begin{enumerate}
\item[(i)]
\begin{equation}
\label{lemma3.1} {\mathcal L}_{\theta,z} \Pi={\mathcal
L}^{(N)}_{\theta,t} \Pi, \end{equation} {\it where index $z$ (resp. $t$)
indicates the action of the CMS operator ${\mathcal L}_{\theta}$
on $z$ (resp. $t$) variables}
\item[(ii)]
\begin{equation}
\label{lemma3.2} \varphi(\Pi) = \prod_{l=1}^N \prod_{i=1}^n
(1-x_{i}t_l)^{-\theta}\prod_{j=1}^m (1-y_{j}t_l)
\end{equation}
\item[(iii)]
\begin{equation}
\label{lemma3.3} \varphi({\mathcal L}_{\theta,z} \Pi)={\mathcal
L}_{n,m,\theta} \varphi(\Pi)
\end{equation}
\end{enumerate}

{\bf Proof.} Introduce the notation
$c_{il}=\frac{z_{i}t_{l}}{1-z_{i}t_{l}}.$ The following identities
are easy to verify $$
\frac{z_{j}c_{il}-z_{i}c_{jl}}{z_{i}-z_{j}}=c_{il}c_{jl}, \quad
\frac{t_{l}c_{ik}-t_{k}c_{il}}{t_{k}-z_{l}}=c_{il}c_{ik}, \quad
z_{i}\frac{\partial}{\partial
z_{i}}(c_{il})=t_{l}\frac{\partial}{\partial
t_{l}}(c_{il})=c_{il}(c_{il}+1). $$ Therefore we have $$
\Pi^{-1}z_{i}\frac{\partial}{\partial z_{i}}(\Pi)=\theta\sum_{l}
c_{il}; \quad\Pi^{-1}t_{l}\frac{\partial}{\partial
t_{l}}(\Pi)=\theta\sum_{i} c_{il}, $$ $$
\Pi^{-1}\left(z_{i}\frac{\partial}{\partial
z_{i}}\right)^2(\Pi)=\theta^2\left(\sum_{l}
c_{il}\right)^2+\theta\sum_{l} c_{il}(c_{il}+1), $$ $$
\Pi^{-1}\left(t_{l}\frac{\partial}{\partial
t_{l}}\right)^2(\Pi)=\theta^2\left(\sum_{i}
c_{il}\right)^2+\theta\sum_{i} c_{il}(c_{il}+1). $$ Therefore $$
\Pi^{-1}{\mathcal L}_{\theta,z}(\Pi)=\theta^2\sum_{i}\left(\sum_{l}
c_{il}\right)^2+\theta\sum_{i,l} c_{il}(c_{il}+1) +2\theta^2
\sum_{i<j}\sum_{l} c_{il}c_{jl} $$ and $$ \Pi^{-1}{\mathcal
L}^{(N)}_{\theta,t}(\Pi)=\theta^2\sum_{l}\left(\sum_{i}
c_{il}\right)^2+\theta\sum_{l,i} c_{il}(c_{il}+1) +2\theta^2
\sum_{k<l}\sum_{i} c_{il}c_{ik},$$ which are equal. This proves
the first formula (\ref{lemma3.1}).

To prove (\ref{lemma3.2}) let us first note that since $\varphi$
is a homomorphism it is enough to consider the case $N=1$ when we
have only one variable $t.$ Introduce now the following automorphism
$\sigma_{\theta}$ which is defined by his action on the power sums as follows:
\begin{equation}
\label{sigma}
\sigma_{\theta}(p_{r}(y)) = -\frac{1}{\theta}
 p_{r}(y).
\end{equation}
Then we have $$ \varphi\left(\prod_{i}
(1-z_{i}t)^{-\theta}\right) = \prod_{i} (1-x_{i}t)^{-\theta}
\sigma_{\theta} \left(\prod_{j} (1-y_{j}t)^{-\theta}\right)= $$ $$
\prod_{i} (1-x_{i}t)^{-\theta}\sigma_{\theta}\left(\exp \log
\prod_{j} (1-y_{j}t)^{-\theta}\right)= $$ $$ \prod_{i}
(1-x_{i}t)^{-\theta}\exp \sigma_{\theta} \left(-\theta\sum_{j\ge
1} \frac{p_{j}(y)t^j}{j}\right) = \prod_{i} (1-x_{i}t)^{-\theta}
\exp \left(-\sum_{j\ge 1} \frac{p_{j}(y)t^j}{j}\right) = $$ $$
\prod_{i}^n (1-x_{i}t)^{-\theta}\prod_{j}^m (1-y_{j}t).$$
This proves the second part of Lemma 3.

To prove the last part introduce
$a_{il}=\frac{x_{i}t_{l}}{1-x_{i}t_{l}},
b_{jl}=\frac{y_{j}t_{l}}{1-y_{j}t_{l}}.$ We have $$
\varphi(\Pi)^{-1}x_{i}\frac{\partial}{\partial
x_{i}}\varphi(\Pi)=\theta\sum_{l} a_{il}, \quad
\varphi(\Pi)^{-1}y_{j}\frac{\partial}{\partial
y_{j}}\varphi(\Pi)=-\sum_{l} b_{jl}, $$ $$
\varphi(\Pi)^{-1}t_{l}\frac{\partial}{\partial
t_{l}}\varphi(\Pi)=\theta\sum_{i} a_{il}-\sum_{j} b_{jl}, $$

$$ \varphi(\Pi)^{-1}\left(x_{i}\frac{\partial}{\partial
x_{i}}\right)^2\varphi(\Pi)=\theta^2\left(\sum_{l} a_{il}\right)^2
+ \theta\left(\sum_{l}a_{il}(a_{il}+1)\right), $$ $$
\varphi(\Pi)^{-1}\left(y_{j}\frac{\partial}{\partial
y_{j}}\right)^2\varphi(\Pi) = \left(\sum_{l} b_{jl}\right)^2 -
\left(\sum_{l}b_{jl}(b_{jl}+1)\right), $$ $$
\varphi(\Pi)^{-1}\left(t_{l}\frac{\partial}{\partial
t_{l}}\right)^2\varphi(\Pi)=\left(\theta\sum_{i} a_{il}-\sum_{j}
b_{jl}\right)^2+\theta\sum_{i}a_{il}(a_{il}+1)-\sum_{l}b_{jl}(b_{jl}+1).
$$ Now taking into account the following identity $$
\frac{(x_{i}+y_{j})(a_{il}-b_{jl})}{x_{i}-y_{j}}=a_{il}+b_{jl}+2a_{il}b_{jl}.
$$ we can write $$ \varphi(\Pi)^{-1}{\mathcal L}_{n,m,\theta}
\varphi(\Pi)=\theta^2\sum_{i=1}^n\left(\sum_{l} a_{il}\right)^2 +
\theta\left(\sum_{i,l}a_{il}(a_{il}+1) -
\sum_{j=1}^m\left(\sum_{l} b_{jl}\right)^2 +
\sum_{j,l}b_{jl}(b_{jl}+1)\right)$$
$$
- \theta\sum_{i=1}^n\sum_{j=1}^m\sum_{l}(a_{il}+b_{jl}+2a_{il}b_{jl})+\theta^2\sum_{1\le
i < j\le n}\sum_{l}(a_{il}+a_{jl}+2a_{il}a_{jl})$$ $$ + \sum_{1\le
i < j\le
m}\sum_{l}(b_{il}+b_{jl}+2b_{il}b_{jl})-(\theta(n-1)-m)\left(\theta\sum_{i,l}
a_{il}-\sum_{j,l} b_{jl}\right) $$ and  $$
\varphi(\Pi)^{-1}{\mathcal
L}^{(N)}_{2,t}\varphi(\Pi)=\sum_{l}\left(\theta\sum_{i} a_{il}-\sum_{j}
b_{jl}\right)^2+\theta\sum_{i,l}a_{il}(a_{il}+1)-\sum_{j,l}b_{jl}(b_{jl}+1)+
$$ $$ 2\theta\sum_{k < l}\left(\theta\sum_{i=1}^n
a_{il}a_{ik}-\sum_{j=1}^m b_{jl}b_{jk}\right).$$ It is easy to
check that the last two expressions are identical, so
$\varphi(\Pi)^{-1}{\mathcal L}_{n,m,\theta} \varphi(\Pi) =
\varphi(\Pi)^{-1}{\mathcal L}^{(N)}_{\theta,t}\varphi(\Pi) = \varphi(\Pi)^{-1}\varphi({\mathcal L}^{(N)}_{\theta,t}(\Pi))=\varphi(\Pi)^{-1}\varphi({\mathcal L}_{\theta,z}(\Pi)).$ This completes the proof of Lemma 3.

To complete the proof of Theorem 1 we should show that the
coefficients $g_{\lambda}(z,\theta)$ in the expansion of the
function $$\Pi = \prod_{i,j} (1-z_{i}t_{j})^{-\theta} =
\sum_{\lambda}g_{\lambda}(z,\theta)m_{\lambda}(t)$$ generate
$\Lambda$ when we increase the number of variables $t.$ But this
follows from the relation (see \cite{Ma}, I.4) $$\sigma_{\theta}
\left(\prod_{i,j} (1-z_{i}t_{j})^{-\theta}\right)=\prod_{i,j}
(1-z_{i}t_{j}) = \sum_{\lambda} (-1)^{|\lambda|} e_{\lambda}(z)
m_{\lambda}(t),$$ where $e_{\lambda}$ and $m_{\lambda}$ are the
standard symmetric functions defined in section 2, and the
fact that $e_{\lambda}$ form a basis in $\Lambda.$ Theorem 1 is
proved.

We believe that a similar statement true for any quantum integral
${\mathcal L}_{f} \in {\mathcal D} (\theta)$ of the CMS problem,
which is equivalent to the fact that the kernel of $\varphi$ is
generated by certain Jack polynomials. For generic values of the parameter $\theta$ this follows from the following Theorem 2.

Let us introduce the set of partitions $H_{n,m},$ which consists of
the partitions $\lambda$ such that $\lambda_{n+1} \leq m$ or, in
other words, whose diagrams are contained in the {\it fat $(n,m)$
- hook} (see fig.1). Its complement we will denote as $\bar
H_{n,m}.$ It consists of the diagrams which contain the $(n+1)\times(m+1)$ rectangle.

\vspace*{1cm}

\begin{center}

\includegraphics[width=7cm]{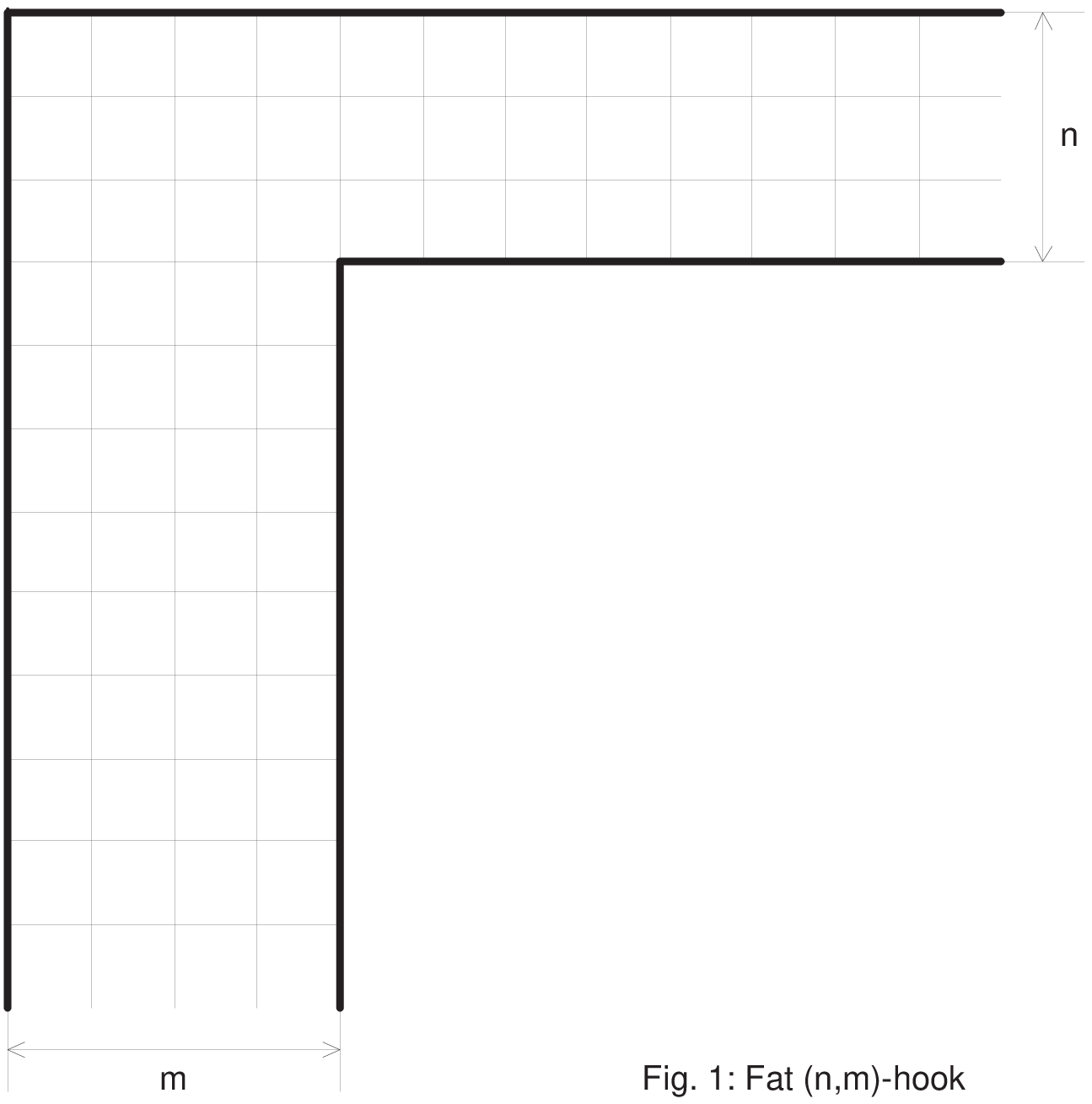}

\end{center}

\vspace*{1cm}

{\bf Theorem 2.}  {\it If $\theta$ is not a negative rational
number or zero, then $Ker \varphi$ is spanned by the Jack
polynomials $P_{\lambda}(z,\theta)$ corresponding to the
partitions which are not contained in the fat $(n,m)$-hook
(or equivalently, which contain $(n+1)\times(m+1)$ rectangle).}

{\bf Proof.} Notice first of all that if $\theta$ is not a
negative rational number or zero the Jack polynomials
$P_{\lambda}(z,\theta)$ are well defined, otherwise in general
they may not exist.

Let us consider the following automorphism (see \cite{Stanley},
\cite{Ma}, VI.10) $\omega_{\theta}$ of algebra $\Lambda:$ $$
\omega_{\theta}(p_{r})=(-1)^{r-1}\theta p_{r}, $$ where $p_r$ are
the standard power sums. Then according to \cite{Ma} (see page
380, formula (10.17)) we have $$
\omega_{\frac{1}{\theta}}\left(P_{\lambda}(z,\theta)\right)=
\theta^{|\lambda|}\frac{H(\lambda,\theta)}{H(\lambda',1/\theta)}P_{\lambda'}(z,1/\theta),
$$ where as before $\lambda'$ is a partition conjugate to
$\lambda.$

Let now $x=(x_{1},x_{2},\dots), y=(y_{1},y_{2},\dots,)$ be two
infinite sequences  variables. Then we have (see \cite{Ma}, page
345, formula $(7.9^{'})$)
\begin{equation}
\label{mac} P_\lambda(x,y,\theta)=\sum_{\mu\subset\lambda}
P_{\lambda/\mu}(x,\theta)P_{\mu}(y,\theta)
\end{equation}
where $P_{\lambda/\mu}(z,\theta)$ are the {\it skew Jack
functions} defined in \cite{Ma} (see Chapter 6, $\S$ 7, 10) and
$\mu\subset\lambda$ means that $\mu_i \leq \lambda_i$ (or
equivalently the diagram of $\mu$ is a subset of the diagram of
$\lambda$).

Now consider the automorphism $\sigma_{\theta}$ which is
combination of the automorphism $\omega_{1/\theta}$ with the
change of sign $y_i \rightarrow -y_i.$ The action of
$\sigma_{\theta}$ on the power sums is given by the formula (\ref{sigma}):
$$
\sigma_{\theta}(p_{r}(y)) = -\frac{1}{\theta}
 p_{r}(y).
$$
If we apply this automorphism acting in $y$
variables on both sides of the formula (\ref{mac}) and put all
the variables $x$ and $y$ except the first $n$ and $m$
of them to zero we get
\begin{equation}
\label{Jack} \varphi
(P_{\lambda}(z,\theta))=\sum_{\mu\subset\lambda}
(-1)^{\mid\mu\mid}P_{{\lambda}
/{\mu}}(x,\theta)\frac{H(\mu,\theta)}
{\theta^{|\mu|}H(\mu^{\prime},1/\theta)}P_{\mu^{\prime}}(y,\theta^{-1}).
\end{equation}

Now let us assume that $\lambda$ is not contained in the fat
$(n,m)$-hook, then $\lambda'_{m+1} > n$. We have two
possibilities: $\mu'_{m+1} > 0$ or $\mu'_{m+1} = 0$. In the first
case we have $P_{\mu}(y_{1},\dots,y_{m},\theta^{-1})=0$ while in the
second case $\lambda'_{m+1} - \mu'_{m+1} > n$, so
according to \cite{Ma} (page 347, formula (7.15)) the skew
function $P_{{\lambda}/{\mu}}(x_{1},\dots,x_{n},\theta)=0$. Thus
we have shown that the Jack polynomials $P_{\lambda}(z,\theta)$
with $\lambda \in \bar H_{n,m}$ belong to the kernel of $\varphi.$

To prove that they actually generate the kernel consider
the image of the Jack polynomials $P_{\lambda}(z,\theta)$ with
$\lambda \in  H_{n,m}.$ From the formula (\ref{Jack}) it follows
that the leading term in lexicographic order of $\varphi
(P_{\lambda}(z,\theta))$ has a form
 $$
( -1 )^{\lambda_{n+1}+\lambda_{n+2}+\dots}x_{1}^{\lambda_{1}}\dots
x_{n}^{\lambda_{1}}y_{1}^{<{\lambda'}_{1}-n>}\dots
  y_{m}^{<{\lambda'}_{m}-n>},
 $$
where $\lambda' =({\lambda'}_{1}, {\lambda'}_{2}, \dots )$ is the
partition conjugate to $\lambda$ and $<a>=\frac{a+|a|}{2} =
\max(0, a).$ From the definition of $\varphi$ it follows that $\varphi
(P_{\lambda}(z,\theta))\in \Lambda_{n,m,\theta}$. It is
clear that all these polynomials corresponding to the diagrams
contained in the fat hook are linearly independent.  Theorem 2
is proved.

Notice that we have also shown that for the generic $\theta$ the
polynomials
\begin{equation}
\label{superJack}
SP_{\lambda}(x,y,\theta)=\varphi(P_{\lambda}(z,\theta)), \quad
\lambda \in H_{n,m}
\end{equation}
form a linear basis in ${\Lambda}_{n,m,\theta}$. These polynomials
are called {\it super-Jack polynomials} (see \cite{Ok}).

{\bf Remark 1.} A natural question is what are the algebraic
generators of the algebra ${\Lambda}_{n,m,\theta}.$ One can show
using Theorem 2 that for generic $\theta$ the first $mn+m+n$
deformed Newton sums (\ref{defNewton})
$p_1(x,y,\theta),\dots,p_{mn+m+n}(x,y,\theta)$ (or equivalently
the super-Jack polynomials $SP_{\lambda}(x,y,\theta)$
corresponding to the Young diagrams with one row of length less or
equal than $mn+m+n$) generate this algebra. The number $mn+m+n$
here is the area of the rectangle $(n+1)\times(m+1)$ minus 1. One
can use Theorem 2 to produce the relations between the generators
(which are the equations determining the generalised
discriminants) but a satisfactory description of them is still
unknown. The fact that ${\Lambda}_{n,m,\theta}$ for generic
$\theta$ is finitely generated was proved earlier in \cite{SV}.

{\bf Remark 2.} For special $\theta$ of the form $- \frac{1}{q}$
the homomorphism $\varphi$ can be passed through the finite
dimenion $N = n+mq$:  $\varphi = \phi \circ \varphi_N,$ where
$\varphi_N : \Lambda \rightarrow \Lambda_N$ is the standard map
(all $z_i$ except $N$ go to zero), and $\phi: \Lambda_N
\rightarrow \Lambda_{n,m,\theta}$ is a homomorphism corresponding
to identification of all $z_i$ except $n$ into $m$ $q$-tuples. The
variety $\tilde {\mathcal D}_{n,m,\theta} =  Spec \,{\mathcal
N}_{n,m,\theta}$  in this case can be interpreted as the stratum
in the discriminant (known also as {\it coincident root locus}) 
consisting of the polynomials of degree $N$
with all but $n$  roots having the same multiplicity $q.$ In
particular, if $m=1$ we have the polynomials with a multiple root
of multiplicity at least $q.$ The corresponding ideal (the kernel
of $\phi$) in this particular case was investigated by B. Feigin,
Jimbo, Miwa and Mukhin in \cite{FJMM}. They showed that it is also spanned by the 
Jack polynomials but the geometry of the corresponding Young diagrams 
is much more complicated. This important paper shows that the case
of special values of $\theta$ is actually very interesting and
deserves more investigation (see \cite{KMSV} for the latest
development in this direction). As we have already mentioned to describe
the algebraic equations of the discriminant strata is a classical problem 
which is still largely open \cite{Cayley, Weyman, Chip}.

{\bf Corollary.}  {\it Let $f \in \Lambda_{\theta}$ be a shifted
symmetric function and ${\mathcal L}^f_{\theta} \in {\mathcal
D}_{\theta}$ be the corresponding quantum integral of the CMS
problem. Then for generic $\theta$ there exists a quantum integral
${\mathcal L}^f_{n,m,\theta}$ of the deformed CMS problem (\ref{defCMS}) such
that the following diagram is commutative} $$\begin{array}{ccc}
\Lambda& \stackrel{{\mathcal
L}^f_{\theta}}{\longrightarrow}&\Lambda
\\ \downarrow \lefteqn{\varphi}& &\downarrow \lefteqn{\varphi}\\
\Lambda_{n,m,\theta}&\stackrel{{\mathcal L}^{f}_{n,m,\theta}}
{\longrightarrow}&\Lambda_{n,m,\theta} \\
\end{array}
$$ {\it The super-Jack polynomials (\ref{superJack}) are the joint
eigenfunctions of all the operators} ${\mathcal L}^f_{n,m,\theta}.$

In the next section we investigate the homomorphism: $f
\rightarrow {\mathcal L}^f_{n,m,\theta}$ in more detail.

\section{Shifted symmetric functions and quantum integrals of the deformed CMS problem}

Let again $P_{n,m}={\mathbb
C}[x_{1},\dots,x_{n},y_{1},\dots,y_{m}]$ be polynomial algebra in
$n+m$ independent variables. The following algebra
$\Lambda_{n,m,\theta}^{\natural}$ introduced in \cite{SV} can be considered
as a shifted version of the algebra $\Lambda_{n,m,\theta}.$ It
consists of the polynomials $p(x_1,\dots,x_n, y_1, \dots, y_m)$
which are symmetric in $x_{1},\dots,x_{n}$ and $y_{1},\dots,y_{m}$
separately and satisfy the conditions $$
f(x_{i}+1/2,y_{j}-1/2)\equiv f(x_{i}-1/2,y_{j}+1/2)$$ on each
hyperplane  $x_{i}+\theta y_{j}=0$ for $i=1,\dots,n$ and
$j=1,\dots,m.$ It is  easy to check that the {\it deformed
Bernoulli sums}
\begin{equation}
\label{defBernoulli}
 b^*_{r}(x,y,\theta)= \sum_{i=1}^n
B_{r}(x_{i}+1/2)+{(-\theta)}^{r-1}\sum_{j=1}^m B_{r}(y_{j}+1/2)
\end{equation}
belong to $\Lambda^{\natural}_{n,m,\theta}$ for all integers $r \geq 0 $.
According to \cite{SV} for generic $\theta$ $b^*_{r}(x,y,\theta)$
generate the algebra $\Lambda_{n,m,\theta}^{\natural}.$

Now we are going to define the homomorphism $\varphi^{\natural},$ which is a
shifted version of the homomorphism $\varphi$ from the previous
section.

Recall that $H_{n,m}$ denote the set of partitions $\lambda$ whose
diagrams are contained in the fat $(n,m)$-hook. Consider the
following {\it Frobenius map} $ F : H_{n,m} \longrightarrow
\mathbb{C}^{n+m}$ : $F
(\lambda)=(p_{1},\dots,p_{n},q_{1},\dots,q_{m}),$ where

\begin{equation}
\label{Frob} p_{i}= \lambda_{i}-\theta
\left(i-\frac{1}{2}\right)-\frac{1}{2}(m-\theta n),\quad
q_{j}=\mu_{j}^{'}-\theta^{-1}
\left(j-\frac{1}{2}\right)+\frac{1}{2}\left(\theta^{-1} m+n\right),
\end{equation}
and $\mu=(\lambda_{n+1},\lambda_{n+2},\dots).$  The motivation for
this particular shift comes from the theory of the deformed root
systems \cite{SV} (see also formula (\ref{rho}) below). The
coordinates $p_i, q_j$ can be considered as a version of the
Frobenius coordinates on the partitions, so we will call them {\it
modified Frobenius $(n,m)$-coordinates} (cf. \cite{IO,VK}).

The image $F(H_{n,m})$ is dense in $\mathbb{C}^{n+m}$ with respect
to Zariski topology. The homomorphism $$ \varphi^{\natural}:
\Lambda_{\theta}\longrightarrow {\bf
C}[x_{1},\dots,x_{n},y_{1},\dots,y_{m}] $$ is defined by the
relation $$ \varphi^{\natural}(f)(p,q)=f((F^{-1}(p,q)),$$ where $(p,q)\in
F(H_{n,m})$. In other words we consider the shifted symmetric
function $f$ as a function on the partitions from the fat hook and
re-write it in the modified Frobenius $(n,m)$-coordinates. The
fact that as a result we will have a polynomial is not obvious. To
prove it consider the following shifted versions of the
Bernoulli sums (\ref{Bernoulli}):
\begin{equation}
\label{defNewton2}
 b^{\natural}_{k}(z,\theta)= \sum_{i\ge1}
\left[B_{k}(z_{i}+\frac{1}{2} +\theta(\frac{1}{2}-i))-B_{k}(\frac{1}{2}+\theta(\frac{1}{2}-i))\right]
\end{equation}

{\bf Lemma 4.} {\it The image $\varphi(f)$ of a shifted symmetric
function $f \in \Lambda_{\theta}$ is a polynomial. For the
Bernoulli sums  $b^{\natural}_{k}(z,\theta)$ it can be given by the
following explicit formula:}
\begin{eqnarray}
\label{frob} \varphi^{\natural}\left(b_{k}^{\natural}(z,\theta)\right)= \sum_{i=1}^n
\left[B_{k}(x_{i}+\frac{1}{2}+\frac{1}{2}(m - \theta
n))-B_{k}(\frac{1}{2}+\theta(\frac{1}{2}-i))\right] \nonumber \\
+ {(-\theta)}^{k-1}\sum_{j=1}^m
\left[B_{k}(y_{i}+\frac{1}{2} - \frac{1}{2}(\theta^{-1} m -
n))-B_{k}(\frac{1}{2}+n+\theta^{-1}(\frac{1}{2}-j))\right]
\end{eqnarray}

{\bf Proof.} Assume that $z_{i}=\lambda_{i}$, where $\lambda\in
H_{n,m}$. Then we have $$ \varphi^{\natural}\left(
b^{\natural}_{k}(\lambda,\theta)\right)=\sum_{i=1}^n\left[B_{k}(\lambda_{i}+\frac{1}{2}+\theta(\frac{1}{2}-i))-B_{k}(\frac{1}{2}+\theta(\frac{1}{2}-i))\right]
$$ $$ + \sum_{ i \ge
1}\left[B_{k}(\mu_{i}+\frac{1}{2}+\theta(\frac{1}{2}-n-i))-B_{k}(\frac{1}{2}+\theta(\frac{1}{2}-n-i))\right]
$$ The same calculations as in Lemma 1 show that the last sum is
equal to $$ {(-\theta)}^{k-1}\sum_{j=1}^{m}
\left[B_{k}(\mu^{\prime}_{j}+\frac{1}{2}+n+\theta^{-1}(\frac{1}{2}-j))-B_{k}(\frac{1}{2}+n+\theta^{-1}(\frac{1}{2}-j))\right].
$$ This proves the formula (\ref{frob}). Since the Bernoulli sums
generate $\Lambda_{\theta}$ this implies the first part of the
Lemma as well.

{\bf Theorem 3.} {\it If $\theta$ is not a negative
rational number or zero, then the image of the homomorphism $\varphi^{\natural}$ coincides
with the algebra ${\Lambda}^{\natural}_{n,m,\theta}$ and the kernel of
$\varphi^{\natural}$ is spanned by the shifted Jack polynomials $
P_{\lambda}^{*}(z,\theta)$ corresponding to the Young diagrams
which are not contained in the fat $(n,m)$-hook.}

{\bf Proof.} The first claim follows from Lemma 4 and Theorem 2
from \cite{SV}. To prove the statement about the kernel consider a
shifted Jack polynomial $P^*_{\lambda}(z,\theta)$ with $\lambda
\in \bar H_{n,m}.$ Let $\mu$ be a partition whose diagram is
contained in the fat $(n,m)$-hook. Since this implies that the
diagram of $\lambda$ is not a subset of the one of $\mu$ according
to the Extra Vanishing Property of shifted Jack polynomials (see
Section 3) we have $P^*_{\lambda}(\mu,\theta)=0.$ Thus we have
shown that $P^*_{\lambda}(z,\theta)$ with $\lambda \in \bar
H_{n,m}$ belong to the kernel of $\varphi.$ To show that they
generate the kernel one should note that $$
 \varphi^{\natural}(P^*_{\lambda}(z,\theta))=\varphi (P_{\lambda}(z,\theta))(x_{1},\dots,x_{n},-\theta y_{1},\dots,-\theta
 y_{m})+\dots,
$$ where dots mean the terms of degree less than $|\lambda|$. From
Theorem 2 it follows that $\varphi^{\natural}(P^*_{\lambda}(z,\theta))$
with $\lambda\in H_{n,m}$ are linearly independent. Theorem is
proved.

{\bf Corollary.} {\it For generic $\theta$ the functions
$$SP^*_{\lambda}(x,y,\theta)=\varphi^{\natural}(P^*_{\lambda}(z,\theta))$$
with $\lambda\in H_{n,m}$ form a basis in
${\Lambda}^{\natural}_{n,m,\theta}$.}

We will call the polynomials $SP^*_{\lambda}(x,y,\theta)$ the {\it
shifted super-Jack polynomials.} They have the following independent definition (cf. \cite{KSa,OkO}
and section 3). 

Let $\lambda$ and $\nu$ be two arbitrary
partitions from the fat $(n,m)$-hook and $(p(\lambda),q(\lambda))$
be the corresponding modified Frobenius $(n,m)$-coordinates
(\ref{Frob}), then the shifted super-Jack polynomials
$SP_{\nu}^{*}$ are uniquely determined by the condition $\deg
SP_{\nu}^{*} \le |\nu|$ and the following property:
\begin{equation}
\label{shsJ}
SP_{\nu}^{*}(p(\lambda),q(\lambda),\theta)=\left\{ \begin{aligned}
\quad H(\lambda),& \quad\quad \quad\lambda=\nu\\
 \quad 0, &\quad \quad |\lambda|\le|\nu|,\lambda\ne\nu
 \end{aligned}
 \right. 
\end{equation}
where $H(\lambda)$ is the same as above (see formula (\ref{H}) in
section 3).

We have also the following version of the {\it Extra Vanishing
Property for the shifted super-Jack polynomials}:
\begin{equation}
\label{EVPsJ} SP^*_{\nu}(p(\lambda),q(\lambda),\theta) = 0
\end{equation}
unless the diagram of $\nu$ is a subset of the diagram of
$\lambda.$

Consider now the algebra of differential operators in $n+m$
variables with rational coefficients belonging to ${\mathbb
C}[x_{1},\dots,x_{n},y_{1},\dots,y_{m}, (x_{i}-x_{j})^{-1},
(x_{i}-y_{l})^{-1},  (y_{k}-x_{l})^{-1}].$ We denote it as $\mathcal{D}(n,m).$

{\bf Theorem 4.} {\it For generic values of $\theta$ there exists
a unique monomorphism $\psi: \Lambda^{\natural}_{n,m,\theta} \rightarrow
\mathcal{D}(n,m)$ such that the following diagram is commutative
$$
\begin{array}{ccc}
\Lambda_{\theta}&\stackrel{\chi}{\longrightarrow}&\mathcal{D}(\theta)
\\ \downarrow \lefteqn{\varphi^{\natural}}& &\downarrow \lefteqn{res}\\
\Lambda^{\natural}_{n,m,\theta}&\stackrel{\psi}{\longrightarrow}&\mathcal{D}(n,m)
\\
\end{array}
$$ where $\chi$ is the inverse Harish-Chandra homomorphism 
and $res$ is the operation of restriction on the generalised
discriminant described in the previous section.}

Indeed let $f$ be a shifted symmetric function from
$\Lambda_{\theta},$ ${\mathcal L}^{f}_{\theta}$ and ${\mathcal
L}^{f}_{n,m,\theta} = res ({\mathcal L}^{f}_{\theta})$ be the
corresponding quantum integrals of CMS and deformed CMS problems.
We know that if $P_{\lambda}(z, \theta)$ is a Jack symmetric
function then  $${\mathcal L}^{f}_{n,m,\theta}
\varphi(P_{\lambda}(z, \theta))=f(\lambda)\varphi(P_{\lambda}(z,
\theta)).$$ Therefore according to Theorem 2 ${\mathcal
L}^{f}_{n,m,\theta}\equiv 0$ if and only if $f(\lambda)=0$ for any
$\lambda$ with the diagram contained in the fat $(n,m)$-hook. Now
from Theorem 3 it follows that $Ker(res \circ \chi) = Ker
\varphi^{\natural}$.

Let us denote by $\mathcal{D}(n,m,\theta)$ the image of
homomorphism $\psi$. We claim that it is generated by the
following quantum integrals of the deformed CMS problem
${\mathcal{L}}_{p}$ introduced in \cite{SV}.

It will be convenient for us to change the notation now by
introducing two sets of indices: $I_{0}=\{1,2,\dots,n\}$ and $
I_{1}=\{\bar1,\bar2,\dots, \bar m\}$ and put $x_{\bar j} = y_j.$
Let $p$ denote the (parity) function on $I = I_{0}\cup I_{1}$ such that
$p(i)=0, i\in I_{0}$ and $p(i)=1, i\in I_{1}$. 

Now define by induction the differential operators $\partial_{i}^{(p)},\quad
i\in I$ as follows: for $p=1$ $$
\partial_{i}^{(1)}=(-\theta)^{p(i)}x_{i}\frac{\partial}{\partial x_{i}}
$$
and for $p > 1$
\begin{equation}
\label{dif1}
\partial_{i}^{(p)}=\partial_{i}^{(1)}\partial_{i}^{(p-1)}-\frac{1}{2}\sum_{j\ne i}{(-\theta)}^{1-p(j)}\frac{x_{i}+x_{j}}{x_{i}-x_{j}}\left(\partial_{i}^{(p-1)}-\partial_{j}^{(p-1)}\right).
\end{equation}
The differential operators ${\mathcal{L}}_{p}$ are defined as the
sum
\begin{equation}
\label{dif2} {\mathcal{L}}_{p}=\sum_{i\in I}{(
-\theta)}^{-p(i)}\partial_{i}^{(p)}.
\end{equation}

{\bf Theorem 5.} {\it The operators  ${\mathcal{L}}_{p}$ for all
$p=1,2,\dots$ map the algebra  $\Lambda_{n,m,\theta} $ into
itself.}

The proof is based on the following two technical lemmas. Let us denote by
$J_{1,\bar1}$ the ideal in the polynomial algebra $\mathbb{C}[x_{1},\dots,x_{n},x_{\bar
1},\dots,x_{\bar m}]$ generated by $x_{1}-x_{\bar 1}$.

{\bf Lemma 5.} {\it The following operators

1) $ \partial_{1}^{(1)}-\theta^{-1}\partial_{\bar1}^{(1)}$

2) $ \partial_{1}^{(1)}+\partial_{\bar1}^{(1)}
-\frac{1}{2}(\theta+1)\frac{x_{1}+x_{\bar1}}{x_{1}-x_{\bar 1}}$

3)
$\partial_{1}^{(1)}\partial_{\bar1}^{(1)}+\frac{1}{2}(-\theta\partial_{1}^{(1)}-\partial_{\bar1}^{(1)})
\frac{x_{1}+x_{\bar1}}{x_{1}-x_{\bar 1}}$

4)
$(\partial_{1}^{(1)})^{2}-\theta^{-1}(\partial_{\bar1}^{(1)})^{2}-(\partial_{1}^{(1)}-\partial_{\bar1}^{(1)})
\frac{x_{1}+x_{\bar1}}{x_{1}-x_{\bar 1}}$

map the ideal $J_{1,\bar 1}$ into itself.}

\noindent The proof is straightforward.

{\bf Lemma 6.} {\it The following operators

1) $\partial_{1}^{(p)}-\partial_{\bar1}^{(p)}$

2) $(\partial_{1}^{(1)}-\partial_{\bar1}^{(1)})\partial_{i}^{(p)}$, $i\ne1,\bar1$

3) $\partial_{1}^{(1)}\partial_{\bar1}^{(p)}-\partial_{\bar
1}^{(1)}\partial_{1}^{(p)}$

4)$(\partial_{1}^{(1)}-\partial_{\bar1}^{(1)})(\partial_{1}^{(p-1)}-\theta^{-1}\partial_{\bar1}^{(p-1)})$
map the algebra $\Lambda_{n,m,\theta} $ into the ideal
$J_{1,\bar1}$. If $f\in \Lambda_{n,m,\theta} $ then
$\partial_{i}^{(p)}f$ is a polynomial.}

Proof is by induction with the use of Lemma 5.

To prove Theorem 5 consider any polynomial $f \in
\Lambda_{n,m,\theta}.$ According to Lemma 6 ${\mathcal{L}}_{p}(f)$
is also a polynomial. To show that it belongs to
$\Lambda_{n,m;\theta}$ we have to check that
$(\partial_{1}^{(1)}-\partial_{\bar1}^{(1)})\left({\mathcal{L}}_{p}\right)(f)$
belongs to the ideal $J_{1,\bar1}.$ But
$$(\partial_{1}^{(1)}-\partial_{\bar1}^{(1)})\left({\mathcal{L}}_{p}\right)(f)=\sum_{i\ne
1,
\bar1}(\partial_{1}^{(1)}+\partial_{\bar1}^{(1)}){(-\theta)}^{-p(i)}\partial_{i}^{(p)}(f)+
(\partial_{1}^{(1)}+\partial_{\bar1}^{(1)})(\partial_{1}^{(p)}-\theta^{-1}\partial_{\bar1}^{(p)})(f),
$$ so due to Lemma 6 this is true since all the summands belong to
$J_{1,\bar1}$. Theorem 5 is proved.

Now according to \cite{SV} the operators ${\mathcal{L}}_{p}$
commute with each other and in particular with
${\mathcal{L}}_{2},$ which is the deformed CMS operator. By the
standard arguments (see e.g. proof of Theorem 1 from \cite{SV})
one can show that they must commute also with the operators
${\mathcal L}^{f}_{n,m,\theta}$ for any $ f \in
\Lambda^{\natural}_{n,m,\theta}.$

Since the operators ${\mathcal L}^{f}_{n,m,\theta}$ separate the
super-Jack polynomials $SP_{\lambda}(x,y,\theta)$ with $\lambda \in
H_{n,m}$ these polynomials must be also the eigenfunctions for the
operators ${\mathcal{L}}_{p}.$  For any $\lambda \in H_{n,m}$ define $\nu_{\lambda} =
(\lambda_{1},\dots,\lambda_{n},\mu_{1}^{\prime},\dots,\mu_{m}^{\prime}),$
where $\mu=(\lambda_{n+1},\lambda_{n+2},\dots).$ It follows from
the results of \cite{SV} that $$
{\mathcal{L}}_{p}SP_{\lambda}(x,y,\theta)={\mathcal{Z}}_{p}(\nu_{\lambda}-\rho)SP_{\lambda}(x,y,\theta)
$$ where
\begin{equation}
\label{rho}
 \rho_{i}=\theta(i-\frac{1}{2})+\frac{1}{2}(m-\theta n),\quad \rho_{n+j}=\theta^{-1}(j-\frac{1}{2})-\frac{1}{2}(\theta^{-1}
m+n),
\end{equation}
$i=1,\dots,n,\,\, j=1,\dots,m$ and $$
{\mathcal{Z}}_{p}(\lambda)=\lambda_{1}^{p}+\dots+\lambda_{n}^{p}+(-\theta)^{p}(\mu_{1}^{p}+\dots+\mu_{m}^{p})+\dots,
$$ where dots mean the terms of less lexicographic order. It was
also proven in \cite{SV} that for generic $\theta$
${\mathcal{Z}}_{p}$ generate the algebra
$\Lambda_{n,m,\theta}^{\natural}$. This means that one can express
any operator ${\mathcal L}^{f}_{n,m,\theta}$ as a polynomial of
the operators ${\mathcal{L}}_{p}.$ Thus we have proved the
following

{\bf Theorem 6.} {\it For generic values of $\theta$ the operators
${\mathcal{L}}_{p}$ for $p=1,2,\dots$ generate the algebra}
$\mathcal{D}(n,m,\theta)$.

\section{ Filters and CMS-invariant ideals in $\Lambda$}

The previous results lead naturally to the following question:
which ideals $I$ in $\Lambda$ are invariant under the action of
CMS-operator $L_{\theta}$ and its quantum integrals ? To answer
this question it is useful to introduce the following notion which
was probably first used by Knop and Sahi \cite{KSa}, but we will
follow here the terminology of the recent paper by Regev
\cite{Reg}.

Let $\mathcal P$ be the set of all partitions ( or Young  diagrams).

{\bf Definition.} The subset $\Omega\subset \mathcal P$ is called
{\it filter} if it is closed under the inclusion. In other words
$\Omega$ is a filter if for any diagram $\lambda\in\Omega$ and any
$\mu$ such that $\lambda\subset\mu$ it follows that
$\mu\in\Omega.$

{\bf Theorem 7.} {\it For generic $\theta$ (more precisely, for $\theta$ not a nonpositive rational) there is a bijection between the set of CMS-invariant ideals in $\Lambda$ and the set of filters.}

{\bf Proof.} CMS-invariance implies that the corresponding ideal  $I$ is a linear span of certain Jack
polynomials $P_{\lambda}=P_{\lambda}(\theta)$:
$$
I = Span( P_{\lambda}(\theta), \lambda\in\Omega_{I})
$$
for some  $ \Omega_{I}\subset\mathcal P$. Let us prove that $ \Omega_{I}$ is a filter. Take any
$\lambda\in \Omega_{I}$ and $\lambda\subset \mu$. We should show that  $P_{\mu}\in I$. When
$\mu-\lambda$ is one box then this follows from Pieri formula \cite{Ma, Stanley}
$$
P_{\lambda}P_{1}=\sum_{\nu}\psi_{\nu/\lambda}(\theta)P_{\nu},
$$
where the sum is taken over partitions $\nu$ obtained by adding one box to $\lambda$. If $j$ is such
that $\nu_{j}=\lambda_{j}+1 $ then $\psi_{\nu/\lambda}(\theta)$ is given by
$$
\psi_{\nu/\lambda}(\theta)=\prod_{i=1}^{j-1}\frac{((j-i-1)\theta+\lambda_{i}-\lambda_{j})((j-i+1)\theta+\lambda_{i}-\lambda_{j}-1)}{((j-
i)\theta+\lambda_{i}-\lambda_{j}-1)((j-
i)\theta+\lambda_{i}-\lambda_{j})}
$$
Since $\theta$ is not a negative rational number or zero all coefficients $\psi_{\nu/\lambda}(\theta)$ are 
nonzero. Now CMS-invariance of $I$ implies that all $ P_{\nu}$ in this sum belong to $I$. By induction
the same is true for all $\nu$, such that $\lambda\subset\nu$.

Now let $\Omega$  be a filter. Let us show that $$
J=Span\left(P_{\lambda}(\theta) ,\lambda\in\Omega\right) $$ is an
ideal in $\Lambda$. It  is enough to show that
$P_{\lambda}e_{r}\in J$, where $e_{r}$ is elementary symmetric
function. But this is a direct consequence of the Pieri formula $$
P_{\lambda}e_{r}=\sum_{\nu}\psi_{\nu/\lambda}(\theta)P_{\nu} $$
where now the sum is taken over partitions $\nu$ such that
$\lambda\subset\nu$ and $\nu-\lambda$ is a vertical $r$-strip.
The coefficients $\psi_{\nu/\lambda}(\theta) $ have the following form (see \cite{Ma}, VI.6): 
$$ \psi_{\nu/\lambda}(\theta) = \prod_{s\in
C_{\nu/\lambda}-R_{\nu/\lambda}}\frac{b_{\nu}(s)}{b_{\lambda}(s)}
$$ where $$
b_{\lambda}(s)=\frac{c^{\prime}_{\theta}+\theta}{c^{\prime}_{\theta}+1}
,\quad {\text if} \quad s\in\lambda,\quad {\text and}\quad 1\quad
{\text otherwise} $$ and $C_{\nu/\lambda}$ (resp.
$R_{\nu/\lambda}$) denote the union of the columns (resp. rows )
of $\nu$ that intersect $\nu/\lambda$ and $c^{\prime}_{\theta}$
are given by the formula (\ref{c'}) above. Note that our assumption on  $\theta$ implies that $c^{\prime}_{\theta}+1 \neq 0.$ Theorem 7 now follows.

We will denote the ideal $I = Span( P_{\lambda}(\theta), \lambda\in\Omega)$ corresponding to filter $\Omega$ as $I_{\Omega}(\theta)$.

Let $\Omega(\lambda^{(1)},\dots,\lambda^{(n)})$ be  the set of
Young diagrams $\mu$ such that $\mu$ contains at least one of the
Young diagrams $\lambda^{(1)},\dots,\lambda^{(n)}.$ It is easy to
see that $\Omega(\lambda^{(1)},\dots,\lambda^{(n)})$ is a filter.
We will call such a filter {\it finitely generated}.

{\bf Lemma 7 (A. Regev \cite{Reg}).}  {\it Any filter is finitely generated.}

This result was proven by Regev  \cite{Reg} using combinatorial arguments. We will give a simpler
algebraic proof using our results from \cite{SV}.

{\bf Proof of Lemma 7.} Any filter contains some rectangle $\pi$.
Consider the ideal $I_{\pi}(\theta)=Span(P_{\lambda}(\theta),
\pi\subset\lambda)$ assuming that $\theta$ is generic. It is
obviously contained in $I_{\Omega}(\theta)$. According to
\cite{SV} for generic $\theta$ the algebra
$\Lambda/I_{\pi}(\theta)$ is finitely generated (see Theorem 5 in
\cite{SV}).

Therefore the ideal $I_{\Omega}(\theta)/I_{\pi}(\theta)\subset\Lambda/I_{\pi}(\theta)$ is finitely generated as well (as an ideal in Noetherian algebra). Let $P_{\lambda^{(1)}}(\theta),\dots, P_{\lambda^{(n)}}(\theta)$ be  Jack polynomials whose images generate $ I_{\Omega}(\theta)/I_{\pi}(\theta)\subset\Lambda/I_{\pi}(\theta)$ . We claim that
$\Omega=\Omega(\pi,\lambda^{(1)},\dots,\lambda^{(n)})$. Let $\lambda\in\Omega$ then
$P_{\lambda}\in I_{\Omega}$. By assumption there exist $f_{1},\dots, f_{n}\in\Lambda$ such that
$$
P_{\lambda}-\sum_{i=1}^n f_{i}P_{\lambda^{(i)}}\in I_{\pi}
$$
Now  using again Pieri formula we conclude that $P_{\lambda}$ is a linear combination of $P_{\nu}$
where $\nu$  contains either $\lambda^{(i)}$ or $\pi$. Lemma is proved.

Let now $\Omega$ be a filter and consider the corresponding algebra $\Lambda_{\Omega}(\theta)=\Lambda/I_{\Omega}(\theta)$.

{\bf Theorem 8.} {\it For any filter $\Omega$ and generic $\theta$ the algebra  $\Lambda_{\Omega}(\theta)$ is finitely generated.}

{\bf Proof.}  Consider a rectangle $\pi\in\Omega$ and the corresponding ideal
$$
I_{\pi}(\theta)=Span(P_{\lambda}(\theta), \lambda\supset\pi)\subset\ I_{\Omega}(\theta)
$$
Since  $ \Lambda_{\pi}(\theta)=\Lambda/I_{\pi}(\theta)$ for generic $\theta$ is finitely generated the algebra $ \Lambda_{\Omega}(\theta)$ is
also finitely generated as a homomorphic image of  $\Lambda_{\pi}(\theta)$. 

Consider now the special case when $\Omega=\Omega(\lambda)$ is
generated by one Young  diagram $\lambda$. If $\lambda$ is $(n+1)\times(m+1)$ rectangle then for generic $\theta$ according to \cite{SV}  $\Lambda_{\lambda}(\theta) = \Lambda_{n,m,\theta}$.
Note that when $m=0$ we have the standard symmetric polynomial algebra $\Lambda_n.$
The question is what happens for non-rectangular Young diagrams.

Recall that an algebra is called a {\it domain} if it has no zero divisors , i. e. $ab=0$ implies that either $a=0$ or $b=0$. An algebra is called {\it nilpotent free} if $a^n=0$ implies $a=0$.

{\bf  Theorem 9.} {\it For generic $\theta$ the algebra $\Lambda_{\Omega(\lambda)}(\theta)$ is nilpotent free. It is a domain if and only if $\lambda$ is a rectangle. The ideal $I_{\Omega(\lambda)}(\theta)$ is the intersection of the prime ideals $I_{\pi_{i}}(\theta),$ corresponding to the maximal rectangular subdiagrams $\pi_{1},\dots,\pi_{k}\subset\lambda.$}

{\bf Proof.} First of all if $\lambda$ is a rectangle $\pi$ this follows from the results of \cite{SV} and Theorem 2 above since the corresponding algebra $\Lambda_{\pi}(\theta)$ can be realised as a subalgebra $\Lambda_{n,m,\theta}$ in the polynomial algebra.  Let now $\lambda$ be arbitrary. Consider all maximal rectangular subdiagrams  $\pi_{1},\dots,\pi_{k}\subset\lambda$, then
 $$
 I_{\Omega(\lambda)}(\theta)=\bigcap_{i=1}^k I_{\pi_{i}}(\theta),
 $$
where each of the ideals $I_{\pi_{i}}(\theta)$ is prime. In particular, $P_{\pi_{i}}(\theta)\notin I_{\Omega(\lambda)}(\theta)$ but the product $ P_{\pi_{1}}(\theta)\dots P_{\pi_{k}}(\theta)\in
I_{\Omega(\lambda)}(\theta).$ Thus  $\Lambda_{\Omega(\lambda)}(\theta)$ contains zero divisors unless $\lambda$ is rectangular.

Show now that it is nilpotent free. Take any $a\in\Lambda$ then if $a^n\in I_{\Omega(\lambda)}(\theta)$ then $a^n\in I_{\pi_{i}}(\theta)$ for all $i=1,\dots,k$. But since $I_{\pi_{i}}(\theta)$ are prime this implies that $a\in I_{\pi_{i}}(\theta)$ and hence $a\in I_{\Omega(\lambda)}(\theta)$, which completes the proof.

{\bf  Corollary.} {\it Affine algebraic variety
 $$
 {\mathcal M}_{\lambda}(\theta)=Spec \, \Lambda_{\Omega(\lambda)}(\theta)
 $$
corresponding to a non-rectangular Young diangram $\lambda$ is reducible. Its irreducible components are the generalised discriminants related to maximal rectangular subdiagrams of $\lambda$.}

Let us discuss now the filters generated by two Young diagrams  $\lambda_{1}$ and $\lambda_{2}$. We
restrict ourselves by the case when  $\lambda_{1}=\pi_{1}$,  $\lambda_{2}=\pi_{2}$  are rectangular.

Let $\pi=\pi_{1}\cap\pi_{2} $ be the intersection of these rectangles. We will assume that $\pi$ is
different from $\pi_{1}$ and  $\pi_{2}$.

{\bf Theorem 10.} {\it If $\pi_{1}$ and  $\pi_{2}$ are two rectangular Young diagrams such that neither of them
contains another then for generic $\theta$ the algebra  $\Lambda_{\Omega(\pi_{1},\pi_{2})}(\theta)$ has nilpotent elements.}

{\bf Proof.} Let the intersection $\pi$ of $\pi_{1}\cap\pi_{2}$ be of size $n\times m$. Consider the
following polynomial
$$
R_{n,m}=\prod_{i,j}(x_{i}-y_{j})^2
$$
It is easy to see that $R_{n,m}\in\Lambda_{n,m,\theta}$ for any $\theta$. One can represent it  as a
sum
$$
R_{n,m}=\sum_{\lambda}c_{\lambda}(\theta) SP_{\lambda}(x,y,\theta)
$$
where $\lambda$ contains the $n\times m$ rectangle. Indeed, under the homomorphism sending
$x_{n}$ and $y_{n}$ to zero  $R_{n,m}$ obviously becomes zero. This means that the corresponding
$P_{\lambda}(\theta)$ belong to the kernel  of
$$
 \varphi_{n-1,m-1} :  \Lambda \longrightarrow\Lambda_{n-1.m-1}
 $$
which is known to consist of $P_{\lambda}(\theta)$ with $\lambda$ containing $n\times m$ rectangle. Consider now the smallest
rectangle $\pi^*$ which contains both $\pi_{1}$ and $\pi_{2}$. If $\pi_{1}$ and $\pi_{2}$ are $n\times
M$ and $N\times m$ rectangles respectively, then $\pi^*$ has the size $N\times M$. We claim that
\begin{equation}
\label{*}
R_{N,M}x_{1}\dots x_{N}=\sum_{\lambda\supset\Pi_{N,M}}c_{\lambda}SP_{\lambda}(x,y,\theta)
\end{equation}
as elements of $\Lambda_{N,M,\theta}$. Indeed
$$
R_{N,M+1}=\sum_{\lambda\supset\Pi_{N,M+1}}c_{\lambda}SP_{\lambda}(x,y,\theta)
$$
in $\Lambda_{N,M+1,\theta}$. Putting $y_{M+1}=0$ we came to (\ref{*}). Now take the natural
homomorphism
$$
\Lambda_{N,M,\theta}\longrightarrow\Lambda_{\Omega (\pi_{1},\pi_{2})}(\theta).
$$
Since the right hand side of (\ref{*}) becomes zero we come to the relation
$$
R_{N,M}x_{1}\dots x_{N}=0
$$
in $\Lambda_{\Omega (\pi_{1},\pi_{2})}(\theta).$ Similarly one has the relation
$$
R_{N,M}y_{1}\dots y_{M}=0
$$
in $\Lambda_{\Omega (\pi_{1},\pi_{2})}(\theta).$ We claim that this implies that  $R_{N,M}^2=0$.
Indeed $R_{N,M}$ is the sum of monomials each of them
is divisible either by the product $x_{1}\dots x_{N}$ or $y_{1}\dots y_{M}$. Theorem 10 is proved.

{\bf Conjecture.} {\it For generic $\theta$ the radical of the ideal
$I_{\Omega (\pi_{1},\pi_{2})}(\theta)$ is equal to $I_{\Omega (\pi)}(\theta)$, where $\pi=\pi_{1}\cap\pi_{2}.$}

Recall that the {\it radical} $R(I)$ of the ideal $I \subset
\Lambda$ consists of the elements $a\in\Lambda$ such that
$a^n\in\Lambda$ for some $n$. It might help to prove this
conjecture if the following {\it Stanley conjecture} about Jack
polynomials would be true: if $$
P_{\lambda}(\theta)P_{\mu}(\theta)=\sum_{\nu}c^{\nu}_{\lambda
\mu}(\theta)P_{\nu}(\theta)$$ is the expansion of the product of
two Jack polynomials then the coefficient $c^{\nu}_{\lambda
\mu}(\theta)\ne 0 $ if and only if $c^{\nu}_{\lambda \mu}(1)\ne 0
$ (see Conjecture 8.4 in \cite{Stanley}). Unfortunately as far as
we know it is still an open question.

Thus we see that rectangular Young diagrams (and
related algebras $\Lambda_{n,m,\theta}$) play a very special role
in the theory of filters and corresponding algebras
$\Lambda_{\Omega}(\theta).$ This shows a fundamental importance of the
generalised discriminants and gives another justification for our
investigation.

\section{Combinatorial formulas}

In this section we give some combinatorial formulas for the
super-Jack polynomials and shifted super-Jack polynomials
generalising the results by Stanley, Okounkov and Olshanski (see
\cite{Stanley,OkO,Ok2}) Let us recall these results.

A tableau $T$ on $\lambda$ is called a {\it reverse tableau} if
its entries strictly decrease down the columns and weakly decrease
in the rows. By $T(\Box)$ denote the entry in the square
$\Box\in\lambda$. The following combinatorial formula for shifted
Jack polynomial was proved by Okounkov in \cite{Ok2}:
\begin{equation}
\label{shiftedjack} P_{\lambda}^{*}(x,\theta)=\sum_{T}
\varphi_{T}(\theta)\prod_{\Box\in\lambda}
\left(x_{T(\Box)}-c_{\theta}(\Box)\right)
\end{equation}
where for a square $\Box = (i,j)$
\begin{equation}
\label{ctheta} 
c_{\theta}(\Box) = (j-1) - \theta (i-1)
\end{equation}
(see formula (2.4) in \cite{OkO}).
Here the sum is taken over all reverse tableaux on $\lambda$ with
entries  in $\{1,2,\dots\}$ and $\varphi_{T}(\theta)$ is the same
weight of tableau as in the combinatorial formula for
ordinary Jack polynomials \cite{Stanley}, \cite{Ma}:
\begin{equation}
\label{jack2} P_{\lambda}(x,\theta)=\sum_{T}
\varphi_{T}(\theta)\prod_{\Box\in\lambda} x_{T(\Box)}.
\end{equation}
We should mention that in \cite{Ma} sum in formula  (\ref{jack2})
is taken over ordinary tableaux but since $P_{\lambda}(x,\theta)$ is
symmetric it also holds if the sum in the
right-hand side is taken over all reverse tableaux. We have also the
following generalisation of (\ref{jack2}) for the skew Jack polynomials
\begin{equation}
\label{jack3} P_{\lambda/\mu}(x,\theta)=\sum_{T}
\varphi_{T}(\theta)\prod_{\Box\in\lambda/\mu} x_{T(\Box)}
\end{equation}
where the sum is taken over all reverse tableaux of shape
$\lambda/\mu$ with entries in $\{1,\dots,\}$ (see \cite{Ma}).

Let us consider now a reverse {\it bitableau} $T$   of type $(n,m)$ and shape
$\lambda$ . We  can view $T$ as a filling of a Young diagram  $\lambda$
by symbols $1<2\dots
<n<1^{\prime}<2^{\prime}\dots< m^{\prime}$ such that its entries
weakly decrease down the columns and  right the rows, besides
entries $1,2\dots ,n$ strictly decrease down the columns and
entries $1^{\prime},2^{\prime}\dots, m^{\prime}$  strictly
decrease in rows. Let $T_{1}$ be a subtableau  in $T$ containing
all symbols $1^{\prime},2^{\prime}\dots, m^{\prime}$ and
$T_{0}=T-T_{1}$.

{\bf Theorem 11.} {\it For generic values of the parameter $\theta$
the super-Jack polynomials can be written as
\begin{equation}
\label{deformedjack}
SP_{\lambda}(x_{1},x_{2},\dots,x_{n},y_{1},y_{2},\dots,y_{m})=\sum_{T}
\varphi_{T}(\theta)\prod_{\Box\in\lambda}x_{T(\Box)}
\end{equation}
where $x_{j^{\prime}}$ is denoted as $y_{j}$ and
 $$
\varphi_{T}(\theta)=(-1)^{|\mu|}
\varphi_{T_{1}^{\prime}}(1/\theta)\varphi_{T_{0}}(\theta)\frac{H(\mu,\theta)}
{\theta^{|\mu|}H(\mu^{\prime},1/\theta)} $$}

Proof follows directly from the formulas (\ref{Jack}),(\ref{jack2}),(\ref{jack3}).

To formulate a similar result for shifted super-Jack polynomials it will be convenient
for us to use instead of the algebra $\Lambda_{n,m,\theta}^{\natural}$ the following algebra $
\Lambda_{n,m,\theta}^{\flat}$ consisting of the polynomials $p(x_1, \dots,
x_n, y_1, \dots, y_m)$ which are symmetric in $x_{i}-\theta(i-1),
i=1,\dots,n$ and $y_{j}-{\theta}^{-1}(j-1), j=1,\dots,m$
separately and satisfy the conditions $$ f(x_{i}+1,y_{j}-1)\equiv
f(x_{i},y_{j}) $$ on each hyperplane $x_{i}+\theta(1-i)=\theta
(y_{j}+n-1)+1-j=0$ for $i=1,\dots,n$ and $j=1,\dots,m.$ It is easy
to check that the shift $\tau$ $$ \tau(x_{i})=
x_{i}+\rho_{i},\quad \tau(y_{j})=\ y_{j}+\rho_{n+j}, $$ $\rho$ is given by the formula
(\ref{rho}), establishes
an isomorphism between $\Lambda_{n,m,\theta}^{\flat}$ and
$\Lambda^{\natural}_{n,m,\theta.}$

Consider the homomorphism ${\varphi}^{\flat}=\tau^{-1}\varphi^{\natural}$
$$ \varphi^{\flat} : \Lambda_{\theta}\longrightarrow
\Lambda^{\flat}_{n,m,\theta} $$

Recall that $H_{n,m}$  is the set
of partitions $\lambda$ such that $\lambda_{n+1}\le m$ and
$\mu=(\lambda_{n+1},\lambda_{n+2},\dots)$. Consider  the following
map $ F^{\flat} : H_{n,m}\longrightarrow \mathbb{C}^{n+m}$ :
$F ^{\flat}(\lambda)=(a_{1},\dots,a_{n},b_{1},\dots,b_{m})$ where
$a_{i}= \lambda_{i}, i=1,\dots,n$ and $b_{j}= \mu^{\prime}_{j},
j=1,\dots,m$ . The set $F^{\flat}(H_{n,m})$ is dense in
$\mathbb{C}^{n+m}$ with respect to Zariski topology. It is easy to
see that $$ \varphi^{\flat}(f)(a,b)=f(F^{\flat -1}(a,b)) $$
where $ (a,b)\in F^{\flat}(H_{n,m})$ and $f \in \Lambda_{\theta}$.

We are going to present a combinatorial formula for the following version of
the shifted super-Jack polynomial $$SP^{\flat} _{\lambda}
(x,y,\theta) =\varphi^{\flat}(P^*_{\lambda}(z,\theta)).$$  Recall
that a reverse tableau $T$ type $(n,m)$ and shape $\lambda$  is a
filling by symbols $ 1<2<\dots < n <1^{'}<2{'}<\dots < m^{'} $
such that

1) in each row (resp. column) of $T$ the symbols decrease  in the weak sense from left to right (resp. from top to bottom)

2) there is at most one marked symbol $j^{'}$ in each row and at most one unmarked symbol $i$ in each column.

By $T(\Box)$ denote the entry in the square  $\Box\in\lambda$.

{\bf Theorem 12.} {\it The following formula is true:
\begin{equation}
\label{combshiftedjack} 
SP^{\flat} _{\lambda} =\sum_{T}
\varphi_{T}(\theta)\prod_{\Box\in\lambda}({(-\theta)}^{p(T(\Box))}x_{T(\Box)}-c_{\theta}(\Box)),
\end{equation}
where $c_{\theta}$ are given by (\ref{ctheta}).}

{\bf Proof.} Let us consider the skew diagram $\lambda/\mu$ and define {\it skew shifted Jack polynomials} by the following natural generalisation of the
Okounkov's formula \ref{shiftedjack}: $$
P_{\lambda/\mu}^{*}(x,\theta)=\sum_{T}
\varphi_{T}(\theta)\prod_{\Box\in\lambda/\mu}
\left(x_{T(\Box)}-c_{\theta}(\Box)\right).$$ In \cite{Ok2}
Okounkov proved that
\begin{equation}
\label{Ok}
P_{\lambda}^{*}(z_{1},z_{2}\dots,\theta)=\sum_{\mu\prec\lambda}\varphi_{\lambda/\mu}(\theta)\prod_{\Box\in\lambda/\mu}
\left(z_{1}-c_{\theta}(\Box)\right)P_{\mu}^{*}(z_{2},z_{3}\dots,\theta),
\end{equation}
where $\mu\prec\lambda$ means
$\lambda_{i+1}\le\mu_{i}\le\lambda_{i}$ and
$\varphi_{\lambda/\mu}(\theta)$ is the same coefficient  as in the formula for
the ordinary Jack polynomials  $$
P_{\lambda}(z_{1},z_{2}\dots,\theta)=\sum_{\mu\prec\lambda}\varphi_{\lambda/\mu}(\theta)
z_{1}^{|\lambda/\mu|}P_{\mu}(z_{2},z_{3}\dots,\theta)$$ 
(see \cite{Stanley} and \cite{Ma}, section VI.10). Applying
the formula (\ref{Ok}) $n$ times we get $$
P_{\lambda}^{*}(z_{1},z_{2}\dots,\theta)=\sum_{\mu\subset\lambda}P_{\lambda/\mu}^{*}(z_{1},z_{2},\dots,z_{n},\theta)P_{\lambda}^{*}(z_{n+1},z_{n+2}\dots,\theta)
$$ and thus $$
\varphi^{\flat}(P_{\lambda}^{*}(z_{1},z_{2}\dots,\theta))=\sum_{\mu\subset\lambda}P_{\lambda/\mu}^{*}(x_{1},x_{2},\dots,x_{n},\theta)\omega^*(P_{\mu}^{*}(z_{n+1},z_{n+2}\dots,\theta)).
$$ Now using the duality (\ref{dual}) we have $$
\varphi^{\flat}(P_{\lambda}^{*}(z_{1},z_{2}\dots,\theta))=\sum_{\mu\subset\lambda}P_{\lambda/\mu}^{*}(x_{1},x_{2},\dots,x_{n},\theta)\frac{H(\mu,\theta)}{H(\mu^{\prime},1/\theta)}P_{\mu^{\prime}}^{*}(y_{1},y_{2}\dots,y_{m},1/\theta).
$$ But according to the formula (\ref{shiftedjack}) $$
P_{\mu^{\prime}}^{*}(y_{1},y_{2}\dots,y_{m},1/\theta)=\sum_{T^{\prime}_{1}}
\varphi_{T^{\prime}_{1}}(1/\theta)\prod_{\Box^{\prime}\in\mu^{\prime}}
\left(x_{T^{\prime}_{1}(\Box^{\prime})}-c_{1/\theta}(\Box^{\prime})\right)=
$$ $$ \sum_{T_{1}} \varphi_{T_{1}}(1/\theta)\prod_{\Box\in\mu}
\left(x_{T_{1}(\Box)}+\frac{1}{\theta}c_{\theta}(\Box)\right)=(-1/\theta)^{|\mu|}\sum_{T_{1}}
\varphi_{T_{1}}(1/\theta)\prod_{\Box\in\mu}
\left((-\theta)x_{T_{1}(\Box)}-c_{\theta}(\Box)\right), $$
where $T^{\prime}_{1}$ is the reverse tableau conjugate to $T_{1}.$
Therefore $\varphi^{\flat}
(P_{\lambda}^{*}(z_{1},z_{2}\dots,\theta))$ can be rewritten as $$ \sum_{T_0}(-1/\theta)^{|\mu|}\frac{H(\mu,\theta)}{H(\mu^{\prime},1/\theta)}
\varphi_{T_0}(\theta)\prod_{\Box\in\lambda/\mu}
\left(x_{T_0(\Box)}-c_{\theta}(\Box)\right) 
\varphi_{T_{1}}(1/\theta)\prod_{\Box\in\mu}
\left((-\theta)x_{T_{1}(\Box)}-c_{\theta}(\Box)\right)$$ $$
= \sum_{T}
\varphi_{T}(\theta)\prod_{\Box\in\lambda}({(-\theta)}^{p(T(\Box))}x_{T(\Box)}-c_{\theta}(\Box)).
$$ This completes the proof.

\section{Some open questions.}

It seems that for the generic values of the parameter $\theta$ the
general situation is more or less clear now, so the main problems remain for the
special values of the parameter $\theta.$ Let us mention some of
them.

For a given negative rational $\theta$ is it true that the kernel
of the homomorphism $\varphi$ is generated by some Jack
polynomials ? If yes, what is the geometry/combinatorics of the
corresponding Young diagrams ? The best results known so far in
this direction can be extracted from the paper \cite{FJMM} (see
also \cite{KMSV}).

How to describe the generators of the algebra
$\Lambda_{n,m,\theta}$ for special values of $\theta$ ? What are
the corresponding Poincare series ? For generic $\theta$ the
answer to the last question was given in \cite{SV}.

How is this related to the theory of quasi-invariants for the deformed root systems \cite{FV} ?
In particular, are there interesting extensions of the algebra
$\Lambda_{n,m,\theta}$ for special values of $\theta$ ? We know
that at least for $m=1$ the answer is positive (see \cite{FV} for
details).

There are also several important open questions left in the case
of generic $\theta.$ In particular, as we have already mentioned
above (see Remark 1 in section 5) the finding of a
convenient set of generators for the algebra
$\Lambda_{n,m,\theta}$ and a satisfactory description of the relations between
them is still to be done. Another problem is to extend our investigation of the
algebras $\Lambda_{\Omega}(\theta)$ to a general filter $\Omega.$

A natural question also is about the generalisations of our
results for the deformed Macdonald-Ruijsenaars operators
introduced in \cite{SV}. We are planning to discuss this in a
separate paper.

\section{Acknowledgements.}

We are grateful to O. Chalykh, M. Feigin, T. Miwa, A. Okounkov, G.
Olshanski and M. Reid for useful and stimulating discussions. We are grateful also to R. Stanley for
his comments on the present situation with one of his conjectures about Jack polynomials.

The second author (A.P.V.) is grateful to IHES (Bures-sur-Yvette,
France) for the hospitality in February 2003, and in particular to
M. Kontsevich, N. Nekrasov and M. Rovinski for helpful
discussions.

This work was partially supported by EPSRC (grant GR/M69548).

\end{document}